\documentclass[fleqn]{mycas-sc}

\usepackage[authoryear,longnamesfirst]{natbib}
\usepackage{amsmath}
\usepackage{pdfpages}
\def\tsc#1{\csdef{#1}{\textsc{\lowercase{#1}}\xspace}}
\tsc{WGM}
\tsc{QE}

\begin{document}
\let\WriteBookmarks\relax
\def\floatpagepagefraction{1}
\def\textpagefraction{.001}

\shorttitle{Axonal diffusivities}    

\shortauthors{Pizzolato {et~al.}}  

\title [mode = title]{Axial and radial axonal diffusivities from single encoding strongly diffusion-weighted MRI}



\author[1,2,3]{Marco Pizzolato}[orcid=0000-0003-2455-4596]
\cormark[1]
\ead{mapiz@dtu.dk pizzolato.marco.research@gmail.com}
\ead[url]{https://sites.google.com/site/pizzolatomarco/}
\credit{Conceptualization of this study, Methodology, Software, Analysis, Data acquisition, Funding, Writing of original draft}
\author[2]{Erick Jorge Canales-Rodr\'{i}guez}[orcid=0000-0001-6421-2633]
\credit{Analysis, Revision of writing}
\author[3]{Mariam Andersson}[orcid=0000-0001-7181-3007]
\credit{Analysis, Revision of writing}
\author[1,3]{Tim B. Dyrby}[orcid=0000-0003-3361-9734]
\credit{Data acquisition, Funding, Analysis, Revision of writing}







\affiliation[1]{organization={Department of Applied Mathematics and Computer Science, Technical University of Denmark},
            city={Kgs. Lyngby},
            country={Denmark}}
\affiliation[2]{organization={Signal Processing Lab (LTS5), \'{E}cole Polytechnique F\'{e}d\'{e}rale de Lausanne},
            city={Lausanne},
            country={Switzerland}}
\affiliation[3]{organization={Danish Research Centre for Magnetic Resonance, Center for Functional and Diagnostic Imaging and Research, Copenhagen University Hospital Amager and Hvidovre},
            city={Copenhagen},
            country={Denmark}}






\cortext[1]{Corresponding author}


\begin{abstract}
We enable the estimation of the per-axon axial diffusivity from single encoding, strongly diffusion-weighted, pulsed gradient spin echo data.
Additionally, we improve the estimation of the per-axon radial diffusivity compared to estimates based on spherical averaging.
The use of strong diffusion weightings in magnetic resonance imaging (MRI) allows to approximate the signal in white matter as the sum of the contributions from axons.
At the same time, spherical averaging leads to a major simplification of the modeling by removing the need to explicitly account for the unknown orientation distribution of axons.
However, the spherically averaged signal acquired at strong diffusion weightings is not sensitive to the axial diffusivity, which cannot therefore be estimated.
After revising existing theory, we introduce a new general method for the estimation of both axonal diffusivities at strong diffusion weightings based on zonal harmonics modeling.
We additionally show how this could lead to estimates that are free from partial volume bias with, for instance, gray matter. 
We test the method on publicly available data from the MGH Adult Diffusion Human Connectome project dataset.
We report reference values of axonal diffusivities based on 34 subjects, and derive estimates of axonal radii.
We address the estimation problem also from the angle of the required data preprocessing, the presence of biases related to modeling assumptions, current limitations, and future possibilities.
\end{abstract}




\maketitle

\section{Introduction}
The \textit{in vivo} characterization of the brain's white matter (WM) tissue with diffusion-weighted magnetic resonance imaging (DW-MRI) requires the exact identification of diffusive properties of its constituents.
However, the formulation of general diffusion biophysical models that explain the signal contributions from all of the tissue constituents at once can lead to challenges in the estimation of the models' unknowns parameters \citep{jelescu2020challenges}.
Great attention has been paid to the strongly diffusion-weighted acquisition regime \citep{jensen2016fiber,veraart2019scaling,ozarslan2018influence}, or high b-value regime, which allows for modeling only the most diffusion-restricted contributions to the diffusion-weighted signal, thus greatly simplifying the modeling task.
In WM, a strong diffusion weighting is broadly defined as when the contribution of the extra-axonal water to the spherically averaged signal can be considered negligible.
The minimum b-value that can be considered as ``high'' ranges between  $4000\,\textrm{s/mm}^2$ \citep{ramanna2020triple} and  $6000\,\textrm{s/mm}^2$  \citep{mckinnon2019measuring} \textit{in vivo}, while a value of $20000\,\textrm{s/mm}^2$ is considered sufficient \textit{ex vivo} \citep{veraart2020noninvasive}.

In WM, a commonly accepted assumption is that the most restricted microstructural constituents that are visible with DW-MRI are axons.
Diffusion within an axon can be modeled as a so-called axisymmetric ``microscopic'' diffusion tensor \citep{kroenke2004nature,lasic15} characterized by two apparent \textit{per-axon} \citep{kaden2016quantitative} diffusivities: one aligned along the main axonal axis, the \textit{axial} or \textit{parallel} diffusivity $\lambda_{\parallel}$, and the other aligned perpendicularly to it, the \textit{radial} or \textit{perpendicular} diffusivity $\lambda_{\perp}$.
To avoid confusion with the axial (AD) and radial (RD) diffusivity values estimated with diffusion tensor imaging (DTI) \citep{basser1994estimation}, we prefer to refer to the \textit{axonal} diffusivities as to the parallel and perpendicular diffusivities.
The axonal diffusivities assume the meaning of average axonal diffusivities when considering an ensemble of axons within an MRI voxel.
In this scenario, \citet{jensen2016fiber} showed that at ``low'' high b-values (e.g., $5000\,\textrm{s/mm}^2$) the spherically averaged signal follows a power law decay which allows to obtain a measurement of $\zeta=f_a/\sqrt{\lambda_{\parallel}}$ that contains the axonal signal fraction $f_a$ and the axonal parallel diffusivity.
Assuming that the parallel diffusivity remains relatively uniform across the WM, $\zeta$ constitutes a good proxy of the axonal signal fraction but remains a non-specific metric. 
Moreover, at higher b-values (e.g., $10000\,\textrm{s/mm}^2$), the spherically averaged signal becomes sensitive to the axonal perpendicular diffusivity $\lambda_{\perp}$ \citep{veraart2019scaling} which determines a deviation from the typical power-law according to an exponential decay.

In the strong diffusion-weighting regime (e.g., $b\ge 4000\,\textrm{s/mm}^2$ \textit{in vivo}), however, the spherically averaged signal, as shown by \citet{andersson2022does} and as we detail later, does not provide a useful way to access the parallel diffusivity as it is insensitive to it.
The parallel diffusivity indeed remains hidden within the observable parameters, such as in $\zeta$.
However, using the spherically averaged signals obtained from a mixture of linear (e.g., single encoding) and prolate (e.g., triple encoding) or planar encodings, \citep{jensen2018characterizing} and \citet{dhital2019intra} could circumvent this issue thus gaining a more direct access to the estimation of the parallel diffusivity.

In this work, we show that both parallel and perpendicular axonal diffusivities can be estimated by considering the full directional signal acquired at strong diffusion weighting with a conventional (single diffusion encoding) pulsed gradient spin echo (PGSE) experiment \citep{stejskal1965spin}.
We demonstrate this on the Human Connectome MGH Adult Diffusion dataset.
In particular, we introduce a method that is general and can, in theory, also account for the presence of bias due to residual partial volume with gray matter or with isotropically restricted compartments \citep{pizzolato2022axonal} such as those arising from cell-nuclei, vacuoles and possibly other types of structures as presented by \citet{andersson2020axon}.
Additionally, the use of single diffusion encoding PGSE offers a higher efficiency in achieving high b-values while minimizing the echo time.
This translates to an overall higher signal-to-noise ratio (SNR), and more efficient separation of the directional contributions from two or more fiber bundles which is useful, for instance, for tractography \citep{rensonnet2020signal}.
At the same time, the PGSE sequence is associated with a clearer definition of physical concepts such as diffusion time, q-value and its relation to the apparent ensemble average propagator \citep{callaghan1988nmr}.

In the following, we first revisit the theory, starting with the isolation of the axonal compartment at strong diffusion weightings, until the estimation of its perpendicular diffusivity through the spherically averaged signal.
We then propose the calculation of the perpendicular diffusivity directly from the ratio of two spherically averaged signal shells.
Afterwards, we propose a method for the estimation of both the parallel and perpendicular diffusivities using the full directional signal from only two single encoding PGSE shells while exploiting relationships across the coefficients of their spherical harmonics expansions.
We will show that this formulation can account also for the presence of isotropic compartments to be insensitive, for instance, to the presence of partial volume with gray matter.
This method is then synthetically validated and tested on the Human Connectome MGH Adult Diffusion Data.

\section{Theory}
The diffusion-weighted signal coming from a voxel can be expressed as the sum of contributions from $J$ different microscopic physicochemical environments, typically referred to as microstructure compartments
\begin{equation}
    S(TE,TR,b,\textbf{u}) \approx \sum_{j=1}^{J} \rho(TE,TR|PD_j,T_{1j},T_{2j}) \Phi_j(b,\textbf{u}|p_j)
\label{eq_compartments}
\end{equation}
where $TE$ is the echo-time, $TR$ the repetition time, $b$ the b-value, $\textbf{u}$ the direction of the diffusion encoding (e.g., the gradient direction), and where $PD_j$, $T_{1j}$, and $T_{2j}$ are the proton density, longitudinal, and transverse relaxation times of the $j$-th compartment.
Note that the representation in Eq.~\ref{eq_compartments} assumes that the relaxation attenuation $\rho(\cdot)$ and the diffusion attenuation $\Phi(\cdot)$ are independent, and that no exchange occurs across compartments: this is the reason for using the approximation symbol, which will be dropped in favor of the equality symbol to make clear when further approximations are made.
In the equation, each $\Phi_j(\cdot)$ corresponds to a different functional form that depends on a generic set of parameters $p_j$.
It is also worth noting that the b-value $b$ is used to loosely indicate diffusion weighting and could be decoupled in the variety of acquisition-specific parameters, such as the pulse duration $\delta$, separation $\Delta$, and gradient strength $g$, although more parameters exist to describe different types of diffusion encoding, e.g., oscillating gradient spin echo (OGSE), b-tensor, and generalized gradient waveforms.
Finally, the function $\rho(\cdot)$ depends only on single longitudinal and transverse relaxation times.
However, depending on the definition of ``compartment'', a distribution of relaxation times might better describe the biophysical complexity found in the tissue \citep{mackay2006insights}. 
In the strong diffusion weighting regime, practically defined as $b\ge b_H$ with $b_H$ being a ``high enough'' b-value, the average along the diffusion gradient directions of the signal corresponding to Eq.~\ref{eq_compartments} mainly contains contributions from the compartments that have ``survived'' the relaxation and diffusion attenuations imparted by the specified $TE$, $TR$, and $b$.
Hence, water within axons and cells, where water diffusion is geometrically restricted or absent, will contribute to the signal measured in WM voxels.
Indeed, the signal contribution from water trapped between the myelin layers disappears because of its short $T_2$ and of the relatively long $TE$ used to reach high b-values.
The major contribution to the signal in the high b-value regime is however determined by axons.
Neglecting for the moment the contributions from non-axonal compartments, additional approximation is required to characterize the signal contributions from axons in the WM tissue.
For instance, it is known that axons within a voxel have different sizes, and modeling of the distribution of radii has been investigated previously \citep{assaf2008axcaliber}.
However, the diffusion-weighted signal is more sensitive to the larger radii and the ``tail'' of the distribution (i.e., larger diameters) is over-represented in the signal \citep{burcaw2015mesoscopic,veraart2020noninvasive}.
Axons, moreover, follow curved trajectories with various degrees of pseudo-randomic tortuousity, whose effects on water diffusion have been quantified by \citet{andersson2020axon}.
It is difficult to account for those effects analytically.
Most importantly, the increased number of parameters to be estimated may lead to degeneracy of the corresponding parameter space \citep{jelescu2016degeneracy}.
Ultimately, the diffusion signal within the axonal compartment is usually represented by diffusion within a cylinder.
In this scenario, and given the typical acquisition parameters used in a single diffusion encoding PGSE sequence, the diffusivity along the direction perpendicular to the cylinder axis may be related to the cylinder's radius using the Gaussian phase approximation \citep{neuman1974spin,vangelderen1994evaluation}
\begin{equation}
    \lambda_{\perp} \approx \frac{2}{\delta^2(\Delta - \delta/3) } \sum_{m=1}^{\infty} \frac{ 2D_0 \alpha_m^2 \delta - 2 + 2e^{-D_0 \alpha_m^2\delta} + 2 e^{-D_0 \alpha_m^2 \Delta} - e^{-D_0 \alpha_m^2 (\Delta - \delta)} - e^{-D_0 \alpha_m^2 (\Delta + \delta)}  }{ D_0^2 \alpha^6\left(R^2 \alpha^2 - 1\right)}
\label{eq_effective_radius}
\end{equation}
where $\delta$ and $\Delta$ are the pulse length and separation, $D_0$ is the intrinsic diffusivity of the medium, and $\alpha_m$ are the roots of the equation $J_1^{'}(\alpha_m R)=0$ with $J_1^{'}$ being the derivative of the Bessel function of the first kind, order one.
Note that, in general, $D_0$ differs from the diffusivity along the direction of the cylinder's axis, $\lambda_{\parallel}$, because of the irregular shape of axons \citep{andersson2020axon,lee2020impact,lee2020impact}.
The axonal directional signal can therefore be related to a (microscopic) axisymmetric diffusion tensor as
\begin{equation}
    \textbf{D}(\textbf{n},\lambda_{\parallel},\lambda_{\perp}) = (\lambda_{\parallel}-\lambda_{\perp})(\textbf{n}\textbf{n}^T) + \lambda_{\perp}\textbf{I}
\label{eq_tensor}
\end{equation}
where $\lambda_{\parallel}$ and $\lambda_{\perp}$ are the parallel and perpendicular diffusivities indicated previously, $\textbf{n}$ is the direction of the cylinder's axis, and $\textbf{I}$ indicates the identity matrix.

\subsection{Orientational invariance}
On a coarse global level, WM is organized in bundles of axons which follow coherent trajectories connecting different brain regions.
Within each bundle, axons are mainly aligned with the bundle's trajectory although they present different degrees of orientational dispersion at different locations along the bundle.
Moreover, two or more bundles with different trajectories can coexist within an MRI voxel.
This is the case of, for instance, the Corticospinal Tract and the Arcuate Fasciculus.
Axons belonging to different bundles can in theory have different biophysical characteristics.
For instance, \citet{girard2017ax} used microstructural features of different bundles identified within a voxel to aid tractography algorithms in recognizing which bundle configuration (e.g., crossing, kissing) was present.
Additionally, \citet{barakovic2021resolving} presented a method where each bundle, represented by a tractography streamline, has its own transverse relaxation time $T_2$.
However, susceptibility effects can lead to directional differences of the measured transverse relaxation time along the bundle trajectory depending on the local angle between the bundle and the direction of the main field in the MRI scanner \citep{sati2012vivo}.
This was confirmed for the axonal compartment by \citet{mckinnon2019measuring} and \citet{pizzolato2022axonal} although attributed to the extra-axonal space by \citet{tax_measuring_2021}.
The presence of these susceptibility dependencies may lead to a scenario where, within a voxel, eventual differences between two or more bundles' transverse relaxation times might be confounded by the additional susceptibility effects or, conversely, perfectly equal bundles can appear to have slightly different $T_2$.
On the other hand, the presence of a wide variety of shapes and structures in the WM tissue and microscopic dispersion, such as undulation, reasonably mitigate the directional differences and homogenize the biophysical characteristics of the microstructure underlying a voxel.
This will be the assumption in this work which is also motivated by the results presented by \citet{christiaens2020need} suggesting that diffusion MRI data in the WM can be described by the convolution of a single orientation distribution with a single biophysical kernel, i.e., the axonal kernel.
The axonal signal can therefore be viewed as
\begin{equation}
    S_a(b,\textbf{u}) = \int_{\textbf{n}\in S^2} p(\textbf{n}) K_a(b,\textbf{u}|\textbf{n},\lambda_{\parallel},\lambda_{\perp}) d \textbf{n}
\label{eq_convolution}
\end{equation}
where $\textbf{u}$ and $\textbf{n}$ are the normalized diffusion gradient and axonal direction, and $p(\textbf{n})$ is the axonal orientation distribution function such that
\begin{equation}
    \int_{\textbf{n}\in S^2} p(\textbf{n}) d\textbf{n} = 1, \, p(\textbf{n})\ge0 \, \forall \textbf{n} \in S^2
\label{eq_fodf_condition}
\end{equation}
and where the axonal kernel is given by
\begin{equation}
    K_a(b,\textbf{u}|\textbf{n},\lambda_{\parallel},\lambda_{\perp}) = e^{-b \textbf{u}^T \textbf{D}(\textbf{n},\lambda_{\parallel},\lambda_{\perp})  \textbf{u}}
\label{eq_kernel}
\end{equation}
with $\textbf{D}(\textbf{n},\lambda_{\parallel},\lambda_{\perp})$ being as specified in Eq.~\ref{eq_tensor}.
By expanding $p(\textbf{n})$ in spherical harmonics, as proposed by \citet{anderson2005measurement}, Eq.~\ref{eq_convolution} can be rewritten as
\begin{equation}
    S_a(b,\textbf{u}) = C \sum_{l=0,even}^{L} \sum_{m=-l}^{l} c^{'}_{lm} 2\pi e^{-b\lambda_{\perp}} \Psi_l\left(b[\lambda_{\parallel}-\lambda_{\perp}]\right) Y_l^m(\textbf{u})
    \label{eq_axonal_signal}
\end{equation}
where $C$ is a constant that accounts for the relaxation and the volume fraction of the axonal compartment, i.e., the volume of the voxel occupied by axons.
The coefficients $c^{'}_{lm}$ are those representing the expansion in spherical harmonics of the orientation distribution function
\begin{equation}
    p(\textbf{u}) = \sum_{l=0,even}^{L} \sum_{m=-l}^{l} c^{'}_{lm} Y_l^m(\textbf{u})
    \label{eq_fodf}
\end{equation}
where the $l$-order subscript only assumes even values, corresponding to even spherical harmonics, because of the antipodal symmetry of the diffusion signal.
A full derivation of the above equations is reported by \citet{anderson2005measurement} and reproposed by \citet{zucchelli2017noddi}.
For convenience, the expressions for the zonal functions $\Psi_l(x)$ (up to $l=12$) are reported in appendix~\ref{sec_zhc} while for the spherical harmonics we use the definition in \citet{descoteaux2007regularized}.

\subsection{Estimation of the perpendicular diffusivity}
\label{sec_perp_diff}
The axonal perpendicular diffusivity can be estimated using the spherically averaged signal acquired in the strong diffusion weighting regime.
In such a regime, as will be clear later, the signal shows little sensitivity to the parallel diffusivity.
This enables approximation, which leads to the calculation of the perpendicular diffusivity directly from data collected from two PGSE shells with different (high) b-values.

\subsubsection{Spherical averaging}
Under the orientational invariance assumption introduced previously, an effective way to avoid modeling the directional signal as in Eq.~\ref{eq_axonal_signal} -- thus obtaining a \textit{rotational invariant} expression for the signal -- is to use the spherically averaged values of the signal acquired on shells collected at different high b-values: for a given b-value, the spherical mean is the value obtained by calculating the average of the signals along all gradient directions.
The expression for the spherically averaged single encoding PGSE signal arising from modeling the axonal kernel as a microscopic axisymmetric tensor \citep{kroenke2004nature} can be derived from Eq.~\ref{eq_axonal_signal} \citep{anderson2005measurement} by noting that the spherical mean of the signal is that corresponding to the term with $l=0$, $m=0$, that is
\begin{equation}
    \bar{S}_a(b) = \frac{C}{2} e^{-b\lambda_{\perp}} \Psi_0\left(b[\lambda_{\parallel}-\lambda_{\perp}]\right)
    \label{eq_smt_zonal}
\end{equation}
where the substitutions $c^{'}_{00}=(4\pi)^{-\frac{1}{2}}$, as result of the equality in Eq.~\ref{eq_fodf_condition}, and $Y_0^0(\textbf{u})=(4\pi)^{-\frac{1}{2}}\, \forall \textbf{u}$ were used.
From this we observe that there are three unknown parameters that need to be estimated for each WM voxel: $C$, $\lambda_{\parallel}$, and $\lambda_{\perp}$.
One way to estimate the unknowns is to collect data from at least three shells with different b-value -- in the strong diffusion weighting regime -- and fit the equation to the corresponding spherical means.
However, the number of parameters can be reduced via approximation.

\subsubsection{Approximation and power law scaling}
By using the definition of the zeroth order zonal function $\Psi_0$ in Eq.~\ref{eq_smt_zonal} it is possible to recognize the more familiar \citep{kroenke2004nature} relation for the spherical mean signal
\begin{equation}
    \bar{S}_a(b) = C e^{-b\lambda_{\perp}} \frac{\sqrt{\pi}}{2} \frac{\textrm{erf}\left[\sqrt{b(\lambda_{\parallel}-\lambda_{\perp})}\right]}{\sqrt{b(\lambda_{\parallel}-\lambda_{\perp})}}
    \label{eq_smt}
\end{equation}
where $\textrm{erf}(x)$ indicates the error function.
By noting that 
\begin{equation}
    \underset{x\to\infty}{lim}\textrm{erf}(x)=1
    \label{eq_erf}
\end{equation}
then, when $b$ and $\lambda_{\parallel}$ are high enough, it is possible to approximate Eq.~\ref{eq_smt} as
\begin{equation}
    \bar{S}_a(b) \approx  \frac{C}{\sqrt{b}}  \sqrt{\frac{\pi}{ 4 (\lambda_{\parallel}-\lambda_{\perp}) } }   e^{-b\lambda_{\perp}}
    \label{eq_smt_approx}
\end{equation}
which was used by \citet{jensen2016fiber} while assuming $\lambda_{\perp}=0$ and implicitly assumed by \citet{veraart2019scaling} to characterize the \textit{power law} behavior of the spherically averaged signal.
The approximation in Eq.~\ref{eq_smt_approx} has less than 1$\%$ error for $b(\lambda_{\parallel}-\lambda_{\perp})\ge3.4$ \citep{jensen2018characterizing}.
By using the power law formalism it is possible to observe that the number of unknown variables can be reduced to two when compared to Eqs.~\ref{eq_smt_zonal} and~\ref{eq_smt}
\begin{equation}
    \bar{S}_a(b) \approx \frac{\beta}{\sqrt{b}} e^{-b\lambda_{\perp}}
\label{eq_power_law_diameter}
\end{equation}
by defining
\begin{equation}
    \beta := C  \sqrt{\frac{\pi}{ 4 (\lambda_{\parallel}-\lambda_{\perp}) } }
\label{eq_power_law_diameter_beta}
\end{equation}
which can be considered as a single composite unknown parameter.
The unknowns $\beta$ and $\lambda_{\perp}$ can therefore be estimated using data from the spherical means of two or more shells collected in the strong diffusion weighting regime.
It is worth noting that such a regime can have different connotations.
While, it is always associated to diffusion weightings that suppress the extra-axonal signal, sometimes it used for supporting modeling assumptions where the perpendicular diffusivity can be considered negligible, e.g., $\lambda_{\perp}\approx0$ \citep{jensen2016fiber,jensen2018characterizing}, which corresponds to the low range of the high b-value regime.
However, in this article the perpendicular diffusivity is always assumed to be non-negligible.

\section{Calculation of axonal diffusivities}
\subsection{Power law ratio formulation}
\label{sec_reducing_unkowns}
We note that the expression in Eq.~\ref{eq_smt} can be simplified by adopting a ratio formalism that considers the signal acquired at two different b-values, $b_1$ and $b_2$, located in the strong diffusion weighting regime.
This eliminates the need for estimating the constant $C$ and only requires fitting $\lambda_{\parallel}$ and $\lambda_{\perp}$.
The ratio signal can be written as
\begin{equation}
    \frac{\bar{S}(b_1)}{\bar{S}(b_2)} = \sqrt{\frac{b_2}{b_1}} e^{(b_2-b_1)\lambda_{\perp}} \frac{\textrm{erf}\left[\sqrt{b_1(\lambda_{\parallel}-\lambda_{\perp})}\right]}{\textrm{erf}\left[\sqrt{b_2(\lambda_{\parallel}-\lambda_{\perp})}\right]}.
    \label{eq_smt_ratio}
\end{equation}
The ratio between error functions, however, will show little sensitivity with respect to $\lambda_{\parallel}-\lambda_{\perp}$ since at high b-values both the error functions in the numerator and denominator will be close to the unity as indicated by Eq.~\ref{eq_erf}.
Therefore, Eq.~\ref{eq_smt_ratio} can be approximated as
\begin{equation}
    \frac{\bar{S}(b_1)}{\bar{S}(b_2)} \approx \sqrt{\frac{b_2}{b_1}} e^{(b_2-b_1)\lambda_{\perp}}
    \label{eq_powlaw_ratio}
\end{equation}
to which we refer to as \textit{power law ratio} (PLR) and which uniquely depends on one single unknown, $\lambda_{\perp}$.
This relationship constitutes an effective method for calculating the perpendicular diffusivity of the intra-axonal compartment by using the spherically averaged signal from two shells with different b-values as
\begin{equation}
    \hat{\lambda}_{\perp} = \textrm{log} \left\{ \frac{\bar{S}(b_1)}{\bar{S}(b_2)} \sqrt{\frac{b_1}{b_2}}  \right\} /(b_2-b_1)
    \label{eq_powlaw_ratio_estimation}
\end{equation}
or, alternatively, the perpendicular diffusivity can be estimated through optimization of Eq.~\ref{eq_powlaw_ratio} using data from two or more shells.

\subsection{Comprehensive estimation of axonal diffusivities}
\label{sec_comprehensive_estimation}
In addition to what discussed in the previous section \ref{sec_reducing_unkowns}, we introduce a framework for the estimation of both the parallel and perpendicular axonal diffusivities from strongly diffusion-weighted MRI single encoding PGSE data.
As opposed to spherical averaging, the framework uses the full directional signal expressed in Eq.~\ref{eq_axonal_signal} from only two shells while still not requiring any modeling assumptions for the axonal orientation distribution function.
In the following, first it will be assumed that the axonal compartment is indeed the only compartment being represented in the strong diffusion weighting regime.
Later, the presence of isotropic compartments is dealt with.

\subsubsection{Assuming that only the axonal compartment is represented in the strongly diffusion-weighted signal}
\begin{figure}[h]
\centering
\includegraphics[width=0.99\textwidth]{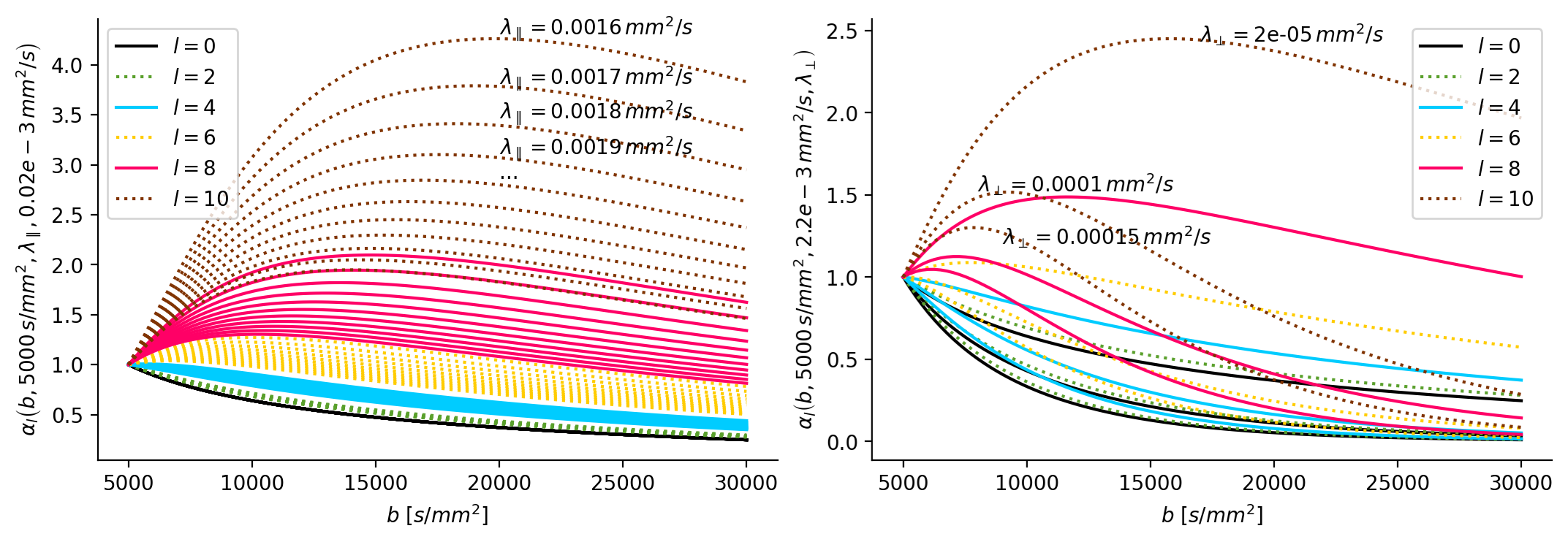}
\caption{Sensitivity of the zonal harmonic ratios $\alpha_l$ to different values of the parallel $\lambda_{\parallel}$ (left) and perpendicular $\lambda_{\perp}$ diffusivity (right) as a function of the b-value of the most diffusion-weighted shell. The values are reported by fixing the lower b-value to $5000\textrm{s}/\textrm{mm}^2$ and varying the other between $5000\textrm{s}/\textrm{mm}^2$ and $30000\textrm{s}/\textrm{mm}^2$. The curves are therefore start at 1 for $5000\textrm{s}/\textrm{mm}^2$.
The left figure reports lines for eleven equidistant values of $\lambda_{\parallel}$ in range $[0.0016,0,0026]\textrm{mm}^2/\textrm{s}$ while the right one illustrates sensitivity for $\lambda_{\perp}\in\{ 0.00002\textrm{mm}^2/\textrm{s},0.0001\textrm{mm}^2/\textrm{s},0.00015\textrm{mm}^2/\textrm{s} \}$. Higher zonal harmonic ratios reveal higher sensitivity to diffusivities.}
\label{fig_senitivities}
\end{figure}

For the sake of explanation, let $b_1$ and $b_2$ be the two b-values used to collect the corresponding diffusion shells in the strong diffusion weighting regime, with $b_1>b_2$.
We begin by expanding the diffusion signal acquired on a shell with b-value $b_2$ up to a truncation order $L$ as proposed by \citet{frank2002characterization,anderson2005measurement,descoteaux2007regularized,hess2006q,canales2009mathematical},
\begin{equation}
    S(b_2,\textbf{n}) = \sum_{l=0,even}^{L} \sum_{m=-l}^{l} c_{lm}(b_2) Y_{l}^{m}(\textbf{n})
\label{eq_sh2}
\end{equation}
where one can obtain estimates for the coefficients $\hat{c}_{lm}(b_2)$.
By comparing  Eq.~\ref{eq_sh2} with Eq.~\ref{eq_axonal_signal} the coefficients of the expansion can be identified as
\begin{equation}
    c_{lm}(b_2) = c^{'}_{lm} C 2\pi e^{-b_2\lambda_{\perp}} \Psi_l\left(b_2[\lambda_{\parallel}-\lambda_{\perp}]\right).
    \label{eq_coefficients_2}
\end{equation}
By expanding the signal acquired for the other b-shell at $b_1$ one can identify a similar relationship for the coefficients
\begin{equation}
    c_{lm}(b_1) = c^{'}_{lm} C 2\pi e^{-b_1\lambda_{\perp}} \Psi_l\left(b_1[\lambda_{\parallel}-\lambda_{\perp}]\right).
    \label{eq_coefficients_1}
\end{equation}
We can then define the \textit{zonal harmonic ratios} for shells at $b_1$ and $b_2$ as
\begin{equation}
\alpha_l(b_1,b_2,\lambda_{\parallel},\lambda_{\perp}) := \frac{c_{lm}(b_1)}{c_{lm}(b_2)} = e^{(b_2-b_1) \lambda_{\perp}} \frac{ \Psi_l\left(b_1[\lambda_{\parallel}-\lambda_{\perp}]\right)}{ \Psi_l\left(b_2[\lambda_{\parallel}-\lambda_{\perp}]\right)}.
\label{eq_coeff_relation}
\end{equation}
It is interesting to note that for $l=0$ and $m=0$ this ratio is approximated by Eq.~\ref{eq_powlaw_ratio}, and thus it is practically insensitive to the parallel diffusivity (at high b-values).
However, this is not the case for coefficients with $l\ge2$ for which the sensitivity to $\lambda_{\parallel}$ increases with $l$ as we show in fig.~\ref{fig_senitivities}.
Notably, Eq.~\ref{eq_coeff_relation} indicates that the ratio between coefficients with the same order $l$ uniquely depends on the axonal diffusivities $\lambda_{\parallel}$ and $\lambda_{\perp}$ and on the chosen b-values.
Most importantly, Eq.~\ref{eq_coeff_relation} allows us to write the spherical harmonics coefficients of one shell as a function of coefficients of the other shell, like
\begin{equation}
    S(b_1,\textbf{n}) = \sum_{l=0,even}^{L} \sum_{m=-l}^{l} \alpha_l(b_1,b_2,\lambda_{\parallel},\lambda_{\perp}) c_{lm}(b_2) Y_{l}^{m}(\textbf{n})
\label{eq_sh1}
\end{equation}
from which we observe that the ratios $\alpha_l(\cdot)$ are the only unknowns if we supply the estimated $\hat{c}_{lm}(b_2)$.
We additionally observe that the ratios only depend on the order $l$.
The expansion in Eq.~\ref{eq_sh1} can therefore be rewritten as
\begin{equation}
    S(b_1,\textbf{n}) = \sum_{l=0,even}^{L}  \alpha_l(b_1,b_2,\lambda_{\parallel},\lambda_{\perp}) \left( \sum_{m=-l}^{l} \hat{c}_{lm}(b_2) Y_{l}^{m}(\textbf{n}) \right)
\label{eq_sh_zonal}
\end{equation}
according to which the $L/2+1$ ratios $\hat{\alpha}_l(\cdot)$ can be estimated linearly.
For a sufficient expansion, i.e., $L\ge2$, it is theoretically possible to non-linearly regress the estimated ratios $\hat{\alpha}_0(\cdot), \hat{\alpha}_2(\cdot),\ldots,\hat{\alpha}_L(\cdot)$,  to optimize for $\lambda_{\parallel}$ and $\lambda_{\perp}$ \citep{Pizzolato2022ismrm}.

\subsubsection{Dealing with the residual presence of isotropic compartments}
\label{ssec_isotropic_removal}
Until now, it has been assumed that only the axonal compartment is represented in the strong diffusion weighting regime.
This is not always the case.
However, fitting a two compartment model at high b-values is prone to degeneracy, further accentuated by the low SNR of the data.
Therefore, we propose to adopt a different strategy.

In order to account for the presence of additional isotropic compartments it is necessary to rewrite Eq.~\ref{eq_axonal_signal} by adding a generic isotropic compartment $\varphi_i(b)$ associated to the zeroth order coefficient -- which only affects the mean of the signal.
This was proposed by \citet{zucchelli2017noddi} to account for the cerebrospinal fluid compartment
\begin{equation}
\begin{split}
    S(TE,b,\textbf{n}) =& c^{'}_{00}\left\{\rho_i v_i e^{-TE/T_{2i}}\sqrt{4\pi}\varphi_i(b) +  \rho_a v_a e^{-TE/T_{2a}} e^{-b\lambda_{\perp}} \sqrt{\pi}\Psi_0\left(b[\lambda_{\parallel}-\lambda_{\perp}]\right) \right\} \\
    +& \sum_{l=2,even}^{L} \sum_{m=-l}^{l} c^{'}_{lm} 2\pi \rho_a  v_a e^{-TE/T_{2a}} e^{-b\lambda_{\perp}} \Psi_l\left(b[\lambda_{\parallel}-\lambda_{\perp}]\right) Y_l^m(\textbf{n})
\end{split}
\label{eq_full_signal}
\end{equation}
where now the dependency on the echo time $TE$ had to be made explicit since the axonal and the isotropic compartments might have different transverse relaxation times, $T_{2i}$ and $T_{2a}$.
Similarly, the volume fractions $v_i$ and $v_a$ now appear (as opposed to the constant $C$) as well as $\rho_i$ and $\rho_a$ which encode the proton densities and the remainders of the longitudinal relaxation for each compartment.
While keeping the echo time constant across the two shells, the same estimation framework presented for Eq.~\ref{eq_sh_zonal} can be used.
However, in this case, the optimization for $\lambda_{\parallel}$ and $\lambda_{\perp}$ needs to be performed only on the zonal ratios $\alpha_l(\cdot)$ for $l\ge2$ which are free from the bias due to the additional isotropic compartment.
This implies that an expansion up to $L=4$ is necessary to have at least two ratios ($l=2$ and $l=4$).
Although only one isotropic compartment is reported in the above equation, this is purely for the sake of simplicity.
In fact, a plurality of isotropic compartments might exist and can be included in the multiplier of $c^{'}_{00}$, each with its volume fraction and relaxation properties.
The proposed method would indeed still apply.
Henceforth, we will refer to an \textit{unbiased} estimator whenever the estimation excludes the signal' spherical mean -- the ratio $\alpha_0$ -- and to a \textit{biased} estimator whenever this is instead included.

\subsection{A regularized variable projection estimation method}
The use of Eq.~\ref{eq_sh_zonal} in the estimation procedure of the axonal diffusivities leads to a linear problem with errors in the variables since the coefficients $\hat{c}_{lm}(b_2)$ may be corrupted by an estimation error.
We then propose an alternative estimation method that mitigates this issue.
This method estimates works with a unique set of coefficients for the two shells and directly leads to the non-linear estimation of the axonal diffusivities.
We define the following linear system
\begin{equation}
\underbrace{
\begin{pmatrix}
S(b_1,\textbf{v}_1)\\
S(b_1,\textbf{v}_2) \\
\vdots\\
S(b_1,\textbf{v}_M)\\
S(b_2,\textbf{w}_1)\\
S(b_2,\textbf{w}_2) \\
\vdots\\
S(b_2,\textbf{w}_N)
\end{pmatrix}}_{y}
=
\underbrace{
\begin{pmatrix}
Y_0^0(\textbf{v}_1) & Y_2^{-2}(\textbf{v}_1) & \cdots & Y_L^{L}(\textbf{v}_1)\\
Y_0^0(\textbf{v}_2) & Y_2^{-2}(\textbf{v}_2) & \cdots & Y_L^{L}(\textbf{v}_2)\\
\vdots  & \vdots  & \ddots & \vdots  \\
Y_0^0(\textbf{v}_M) & Y_2^{-2}(\textbf{v}_M) & \cdots & Y_L^{L}(\textbf{v}_M)\\
\alpha_0(b_2,b_1,\lambda_{\parallel},\lambda_{\perp})Y_0^0(\textbf{w}_1) & \alpha_2(b_2,b_1,\lambda_{\parallel},\lambda_{\perp})Y_2^{-2}(\textbf{w}_1) & \cdots & \alpha_L(b_2,b_1,\lambda_{\parallel},\lambda_{\perp})Y_L^{L}(\textbf{w}_1) \\
\alpha_0(b_2,b_1,\lambda_{\parallel},\lambda_{\perp})Y_0^0(\textbf{w}_2) & \alpha_2(b_2,b_1,\lambda_{\parallel},\lambda_{\perp})Y_2^{-2}(\textbf{w}_2) & \cdots & \alpha_L(b_2,b_1,\lambda_{\parallel},\lambda_{\perp})Y_L^{L}(\textbf{w}_2) \\
\vdots  & \vdots  & \ddots & \vdots  \\
\alpha_0(b_2,b_1,\lambda_{\parallel},\lambda_{\perp})Y_0^0(\textbf{w}_N) & \alpha_2(b_2,b_1,\lambda_{\parallel},\lambda_{\perp})Y_2^{-2}(\textbf{w}_N) & \cdots & \alpha_L(b_2,b_1,\lambda_{\parallel},\lambda_{\perp})Y_L^{L}(\textbf{w}_N)
\end{pmatrix}}_{G\left(\lambda_{\parallel},\lambda_{\perp}\right)}
\underbrace{
\begin{pmatrix}
c_{00}\\
c_{2-2}\\
\vdots\\
c_{LL}
\end{pmatrix}}_{c}
\label{eq_lin_sys}
\end{equation}
where $\textbf{v}_i$ and $\textbf{w}_j$ are one of the $M$ and $N$ gradient directions of shells $b_1$ and $b_2$ respectively, so that $y$ has size $M+N$, $G$ is a matrix of size $(M+N)\textrm{x}(L+1)(L+2)/2$, and $c$ is the vector of coefficients of size $(L+1)(L+2)/2$.
We solve the system in a regularized variable projection (VP) fashion as
\begin{equation}
    \hat{\lambda}_{\parallel},\hat{\lambda}_{\perp} = \underset{\lambda_{\parallel},\lambda_{\perp}}{\operatorname{argmin}} \left\Vert y-G\left(\lambda_{\parallel},\lambda_{\perp}\right)  \left[ G\left(\lambda_{\parallel},\lambda_{\perp}\right)^TG\left(\lambda_{\parallel},\lambda_{\perp}\right)+\gamma R  \right]^{-1} G\left(\lambda_{\parallel},\lambda_{\perp}\right)^T y \right\Vert 
\label{eq_optimization}
\end{equation}
where $R$ indicates either the Laplace-Beltrami (LB) regularization matrix as defined by \citet{descoteaux2007regularized} or the identity matrix \citep{hess2006q} to perform Tikhonov (TK) regularization. The parameter $\gamma$ determines the amount of regularization.
The optimization is carried out iteratively.
When considering the presence of residual isotropic compartments it will be sufficient to remove the mean from each of the two shells composing the vector $y$ and remove the column of $G\left(\lambda_{\parallel},\lambda_{\perp}\right)$ corresponding to the order $l=0$, i.e., the first one.
Note that the linear system formulation in Eq.~\ref{eq_lin_sys} can be expanded to account for more than two shells in a straightforward manner.

\section{Materials and methods}

\subsection{Human data}

Data from 34 subjects of the Human Connectome Project (HCP) Adult Diffusion database collected at the Massachusetts General Hospital (MGH) was used for the study \citep{setsompop2013pushing}.
The single encoding diffusion-weighted images (DWIs) were acquired using a  monopolar pulse (PGSE) Spin-echo EPI sequence and are organized into 40 $b=0$ volumes plus four shells: 64 directions for $b=1000\,\textrm{s/mm}^2$ and for $b=3000\,\textrm{s/mm}^2$, 128 directions for $b=5000\,\textrm{s/mm}^2$, and 256 directions for $b=10000\,\textrm{s/mm}^2$.
Other parameters are the repetition time TR$=8.8\textrm{s}$, echo time TE$=57\textrm{ms}$, gradient pulse duration $\delta=12.9\textrm{ms}$ and separation $\Delta=21.8\textrm{ms}$, field of view FOV$=210\textrm{mmx}210\textrm{mm}$, acquisition matrix of $140\textrm{x}140$ voxels, for 96 slices yielding a $1.5\textrm{mm}$ isotropic voxel resolution, iPAT=3, and partial fourier 6/8.
Full description of the parameters is available on the web\footnote{\url{https://www.humanconnectome.org/study/hcp-young-adult/document/mgh-adult-diffusion-data-acquisition-details}}.
The data was acquired on a Siemens 3T MAGNETOM ConnectomA syngo MR D11 (Siemens Heathineers, Erlangen, Germany) capable of reaching a gradient strength of about 300$\textrm{mT/m}$.

\subsection{Data preprocessing}
The data comes with basic preprocessing that corrects for gradient non-linearities.
We denoised the data using the method described by \citet{ma2020denoise}.
This denoising is based on the estimation of the noise variance in association with a Rician variance stabilization transformation (VST) \citep{foi2011noise}, after which optimal shrinkage with respect to the mean squared error \citep{gavish2017optimal} is applied to the singular values extracted from local 3D cubic patches of data, i.e., local principal component analysis (PCA).
The denoising removes the Rician bias and increases the SNR of the images.
Since the estimation of the noise variance based on PCA \citep{veraart2016diffusion,cordero2019complex} -- as implemented in MRtrix3\footnote{\url{https://www.mrtrix.org/}} -- revealed unstable on the used data, it was replaced with a different in-house method inspired by \citet{gasser1986residual}. 
After denoising, Gibbs ringing removal according to \citet{kellner2016gibbs} was applied using the implementation available in MRtrix3.
Motion artifacts were then corrected using FSL's eddy \citep{andersson2016integrated}.
To further improve the performance of the image preprocessing pipeline and reduce the variability of estimates obtained from the unbiased estimator, we decided to optionally adopt a self-supervised machine learning method, \textit{patch2self} \citep{fadnavis2020patch2self}, as implemented in Dipy\footnote{\url{https://dipy.org/}} \citep{garyfallidis2014dipy} using standard parameters.
The rationale behind using this method is that it can be applied at any point of the processing pipeline and that it is expected to reduce the variance of the estimates by suppressing the remaining residual noise after denoising.
Synthetic tests on the raw DWIs revealed however a poorer performance compared to the adopted denoising based PCA optimal shrinkage (even when including a prior VST), which motivated our choice for the latter as a main denoising step.
Nevertheless, when used as an additional step to a previous denoising, it improved the interpretability of the obtained diffusivity maps and will therefore be considered in the analysis.

\subsection{Synthetic data}
\label{ssec_synthetic_data}
The synthetic data was generated based on the acquisition protocol for the HCP data, with diffusion parameters set to values in line with \textit{in vivo} acquisition.
To simulate the data we first estimated the tissue signal fractions for white matter (WM), gray matter (GM), and cerebrospinal fluid (CSF) using the multi-shell multi-tissue constrained spherical deconvolution framework \citep{jeurissen2014multi} in its generalized version implemented in the Dmipy\footnote{\url{https://github.com/AthenaEPI/dmipy}} library \citep{fick2019dmipy}.
The fiber orientation distribution function (fODF) associated to the WM compartment was estimated with spherical harmonics up to order $L=8$.
The signal fractions were used to regenerate the synthetic brain dataset.
For WM, two compartments were simulated: an axonal compartment with volume fraction of 0.7 and an extra-axonal compartment with volume fraction 0.3.
The axonal and extra-axonal compartments were simulated with axisymmetric tensors having parallel and perpendicular diffusivities of 2.2e-9$\textrm{m}^2/\textrm{s}$ and 2e-11$\textrm{m}^2/\textrm{s}$, and of 1.5e-9$\textrm{m}^2/\textrm{s}$ and 1.04e-9$\textrm{m}^2/\textrm{s}$ respectively.
The GM and CSF compartments were simulated as Gaussian isotropic diffusion components with diffusivities 0.9e-9$\textrm{m}^2/\textrm{s}$ and 3e-9$\textrm{m}^2/\textrm{s}$ respectively.
The transverse relaxation times were fixed to $60\textrm{ms}$ for both the axonal and extra-axonal compartments, to $80\textrm{ms}$ for GM, and to $2\textrm{s}$ for CSF.
Rician noise was applied to the ground-truth data with SNR 20.
This was defined as the ratio between the median value of the intensity of the $b=0$ volume over the white matter (WM fraction above 0.45) and the standard deviation of the additive complex Gaussian noise from which the Rician volumes are generated.
The synthetic datasets were then denoised using the same strategy used for the HCP data, while however providing the correct value for the noise variance.
No further processing was performed except for the additional use of patch2self when specified.

\subsection{Modeling, optimization, and visualization}
The proposed methods are applied on the shells at $b=5000$$\textrm{s/mm}^2$ and $b=10000$$\textrm{s/mm}^2$.
The code has been implemented in Python\footnote{\url{https://www.python.org/}}.
The evaluation of Eq.~\ref{eq_optimization} was carried out with NumPy\footnote{\url{https://numpy.org/}} \citep{2020NumPy-Array} and SciPy\footnote{\url{https://www.scipy.org/}} \citep{2020SciPy-NMeth} using spherical harmonics order (SHO) $L=10$ and $L=12$.
The estimates were obtained with the L-BFGS-B algorithm \citep{byrd1995limited} implemented in SciPy using the following parameters: 'maxcor=20', 'ftol=2.220446049250313e-13', 'gtol=1e-11', 'eps=1e-13', 'maxfun=15000', 'maxiter=15000', and 'maxls=20'.
The admitted ranges for the estimates of parallel and perpendicular diffusivity are  $\lambda_{\parallel}\in[0.0012,0.0034]\textrm{mm}^2/\textrm{s}$ and $\lambda_{\perp}\in[0.000001,0.0002]\textrm{mm}^2/\textrm{s}$.
The initial guess for the estimates were set to the midpoints of the ranges.
Using an analogous optimization setup we estimated the MR radius, $R$, from Eq.~\ref{eq_effective_radius} (truncating the summation to 100 terms), setting the admissible range as $R\in[0,7]\mu\textrm{m}$. 
The visualization of the results was performed with Matplotlib\footnote{\url{https://matplotlib.org/}} \citep{Hunter_2007}.

\subsection{Human data analysis}
We have processed data from 34 subjects of the MGH-HCP Adult Diffusion dataset.
By using MRtrix3 \citep{tournier2019mrtrix3} we created a template based on the WM fiber orientation distribution obtained from multi-shell multi-tissue deconvolution \citep{jeurissen2014multi,dhollander2019improved} using the full acquisition scheme.
After warping each subject's WM fraction into the template space we calculated its mean and standard deviation across subjects for each voxel, and accepted as part of the WM mask all the voxels with a mean larger than 0.25 and a standard deviation lower than 0.09.
For each subject (in native space) we obtained different versions of estimates of $\lambda_{\parallel}$, $\lambda_{\perp}$, and MR radius ($R$).
For comparison purposes, we calculated the axial (AD) and radial (RD) diffusivity from the diffusion tensor fitted to the overall data while considering the correction for Kurtosis \citep{jensen2005diffusional,lu2006three} using MRtrix3.
Although there may be concerns on the validity of this modeling for very high b-values, the estimated parameter maps of AD and RD are normal-appearing.
We then warped each subject's maps into template space.

We performed an in-depth analysis of the corpus callosum by subdividing it into different regions of interest (ROIs).
We calculated the intraclass correlation coefficient ICC2k (average random raters or two-way random effects average of k=34 raters).
This metric assumes values between 0 and 1 and a value above 0.75 is considered as good correlation \citep{koo2016guideline}. The selected metric, in particular, increases with the agreement of the raters -- in this case the various subjects -- in assessing the average value of the diffusivity or MR radius in each ROI, while assuming that the subjects are just a selection of subjects among a much larger population.

\section{Results}
\subsection{Synthetic data}
\begin{figure}[h]
\centering
\includegraphics[width=0.97\textwidth]{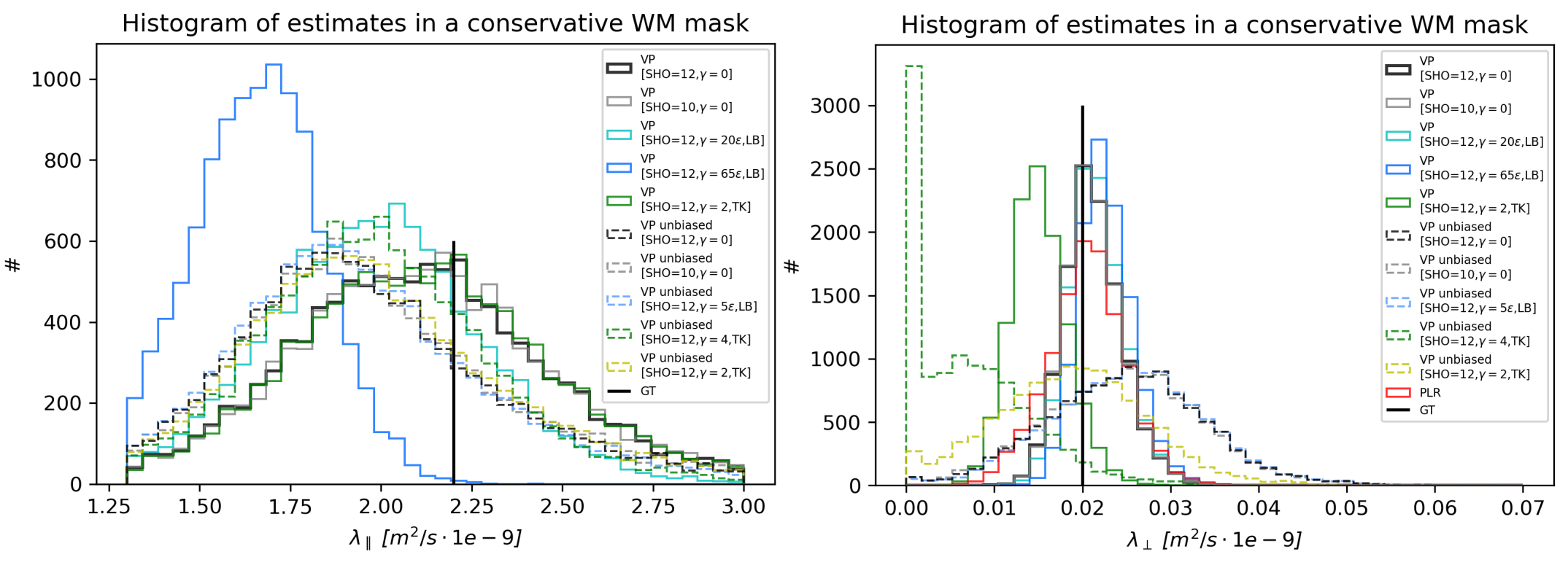}
\caption{Synthetic data, SNR=20. Histograms of estimates obtained with the variable projection (VP) method using different types and amount of regularization and different estimators (including or excluding the signal' spherical mean, i.e., biased or unbiased) for the parallel (left) and perpendicular (right) axonal diffusivity. On the y axis we report the number of voxels. Black lines indicate the ground-truth (GT) values. $\epsilon^{-1}=12000$. In red (right), the results for the non-linear optimization of Eq.~\ref{eq_powlaw_ratio} with the power law ratio (PLR) method.}
\label{fig_synth_hist}
\end{figure}

\begin{figure}[h]
\centering
\includegraphics[width=0.97\textwidth]{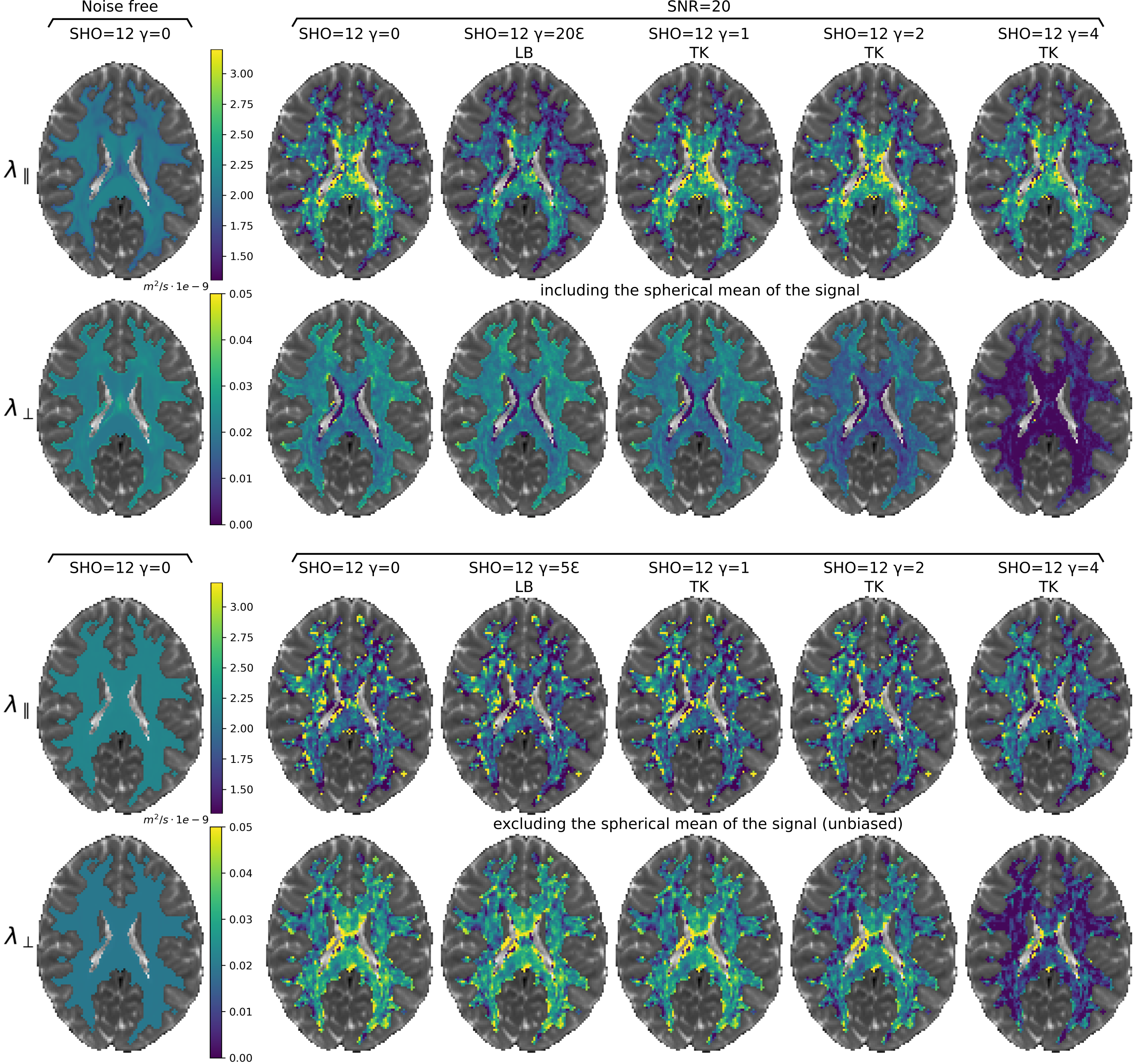}
\caption{Synthetic data, SNR=20. Maps of the obtained axonal diffusivities with the variable projection method for different regularization types, amounts, and using different estimators (biased, unbiased). Maps obtained in the absence of noise are also reported. SHO indicates the spherical harmonics order used. $\epsilon^{-1}=12000$.}
\label{fig_synth_maps}
\end{figure}

Figures~\ref{fig_synth_hist} and~\ref{fig_synth_maps} illustrate the histograms and the maps corresponding to the estimates of parallel and perpendicular diffusivity obtained with the variable projection method for the synthetic data.
Generally, \textit{unbiased} estimates (that do not consider the spherical mean of the signal) for $\lambda_{\perp}$ have a larger variance (less robustness) and can lead to more bias compared to estimates that consider instead the signal spherical mean (mode of estimates shifted compared to the ground-truth).
This can partially be mitigated using regularization.
In general, regularization reduces the variability of the estimates at the cost of more introduced bias.
For unbiased estimates, Tikhonov (TK) regularization (e.g., with $\gamma=2$) seems to be more effective while Laplace-Beltrami (LB) regularization (e.g., with $\gamma=20\epsilon$, $\epsilon^{-1}=12000$) is more effective otherwise.
This assessment is based on considering the closeness between the mode of the distributions and the corresponding ground-truth values, as well as by visually inspecting the maps in fig.~\ref{fig_synth_maps}.
For the case of estimates that consider the spherical mean, the maps look smoother when using LB regularization with $\gamma\approx20\epsilon$, as otherwise many WM areas present high variability and overestimation of the parallel diffusivity.
The reduction of SHO from 12 to 10 does not significantly affect the estimates.
A stronger reduction could in principle improve stability at the cost of reduced sensitivity to changes in the parallel diffusivity.
For this reason, we prefer to use a high SHO and then stabilize the estimation through regularization.

\subsubsection{Bias due to residual partial volume with gray matter}
\begin{figure}[h]
\centering
\includegraphics[width=0.97\textwidth]{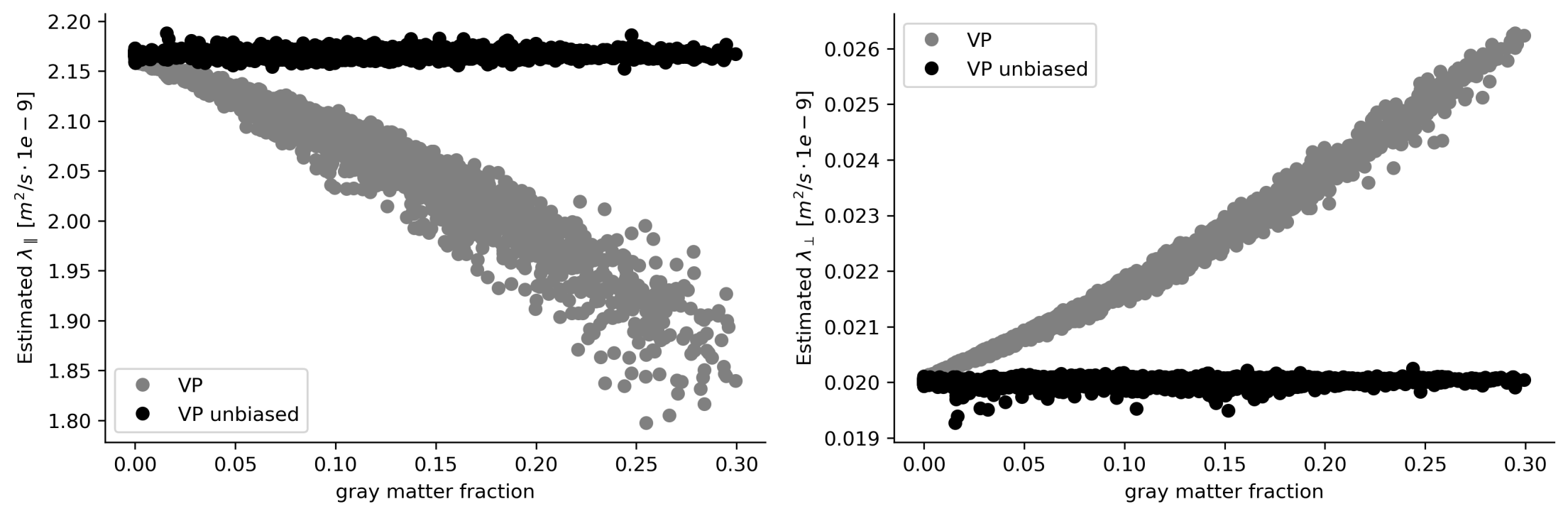}
\caption{Synthetic data. Bias in the axonal parallel (left) and perpendicular (right) diffusivities in noise-free data with the variable projection (VP) method when including (gray) or excluding (black) the signal' spherical mean.}
\label{fig_bias}
\end{figure}

While the noise-free maps in fig.~\ref{fig_synth_maps} for the unbiased estimator look perfectly flat, indicating a correct estimation of the simulated parallel and perpendicular axonal diffusivities (which were set to a constant value across the brain), this is not the case for the estimates obtained when considering also the spherical mean of the signal.
Using the simulated data, therefore, we could investigate this behavior and observe the presence of a correlation between the fraction of gray matter within a voxel and the bias of the estimated diffusivities, fig.~\ref{fig_bias}.
When considering estimates that use the spherical mean (biased) there is an underestimation of the parallel diffusivity up to about 20$\%$  and an overestimation of the perpendicular diffusivity up to about 35$\%$.
The magnitude of the misestimation is however linked to the choice of the diffusive properties in our simulation setup and are therefore expected to change in different scenarios, e.g., the bias may reduce with higher diffusivity or $T_2$ of gray matter.
The results therefore indicate that in the strongly diffusion weighted regime there will be a bias due to partial volume with gray matter, unless the estimates are computed without considering the spherical mean of the signal.
In the noise-free case, differences between the unbiased estimates and the nominal simulated values (in the order of approximately 2$\%$), i.e., 2.2e-9$\textrm{m}^2/\textrm{s}$ and 0.02e-9$\textrm{m}^2/\textrm{s}$, have to be attributed to the fact that the method is implementing an approximation -- by only considering the axonal compartment thus excluding the extra-axonal one.

\subsubsection{Bias due to axonal dispersion and multiple bundles}
\label{ssec_dispersion}
\begin{figure}[h]
\centering
\includegraphics[width=0.97\textwidth]{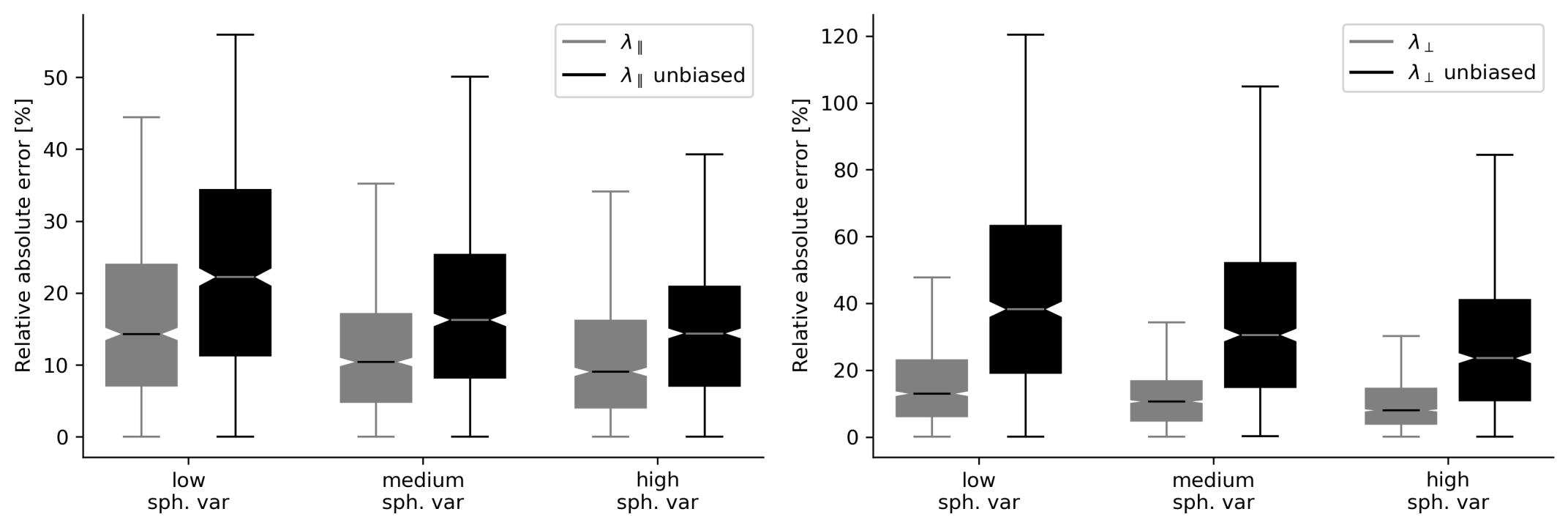}
\caption{Synthetic data, SNR=20. Relative absolute error on the estimated axonal parallel (left) and perpendicular (right) diffusivity obtained when using the VP method when considering the signal' spherical mean (gray) or excluding it (black).}
\label{fig_spherical_variance}
\end{figure}

The proposed method substantially relies on the estimation of spherical harmonics coefficients from the data, from which the zonal coefficients of the axonal kernel are indirectly derived.
By considering the case of a signal within a voxel that is perfectly isotropic (on both shells), then its representation in spherical harmonics will lead to null coefficients except for the first one, i.e., $c_{00}$ which is directly proportional to the spherical mean of the signal.
In this case, the estimation of the axonal diffusivities without considering the spherical mean of the signal (unbiased estimates) would be unfeasible.
While it is safe to assume that such a scenario is unrealistic in WM, where some other spherical harmonics coefficients will be non-zero, the effect of noise will cause more error whenever the values of these coefficients is low.
To assess this, we computed the spherical variance of the FOD used to simulate the data in all WM voxels, similarly to what done in \citet{pizzolato2022axonal}.
This metric is defined as the sum of the squares of the spherical harmonics coefficients  with zonal order $l>0$ \citep{zucchelli2020computational}.
Intuitively a zero spherical variance corresponds to the case of isotropic diffusion, while maximum variance is obtained for a very anisotropic signal such as that arising from diffusion within a stick, i.e., the spherical variance is correlated to fractional anisotropy.
At intermediate values we find signals corresponding to voxels where axons display orientational dispersion and/or crossings.
In fig.~\ref{fig_spherical_variance} we illustrate the relative absolute error as a function of the level of ground-truth spherical variance classified as low, medium, and high.
The classification has been carried out by equally distributing the number of WM voxels falling into each category.
The results confirm that as the orientation distribution function differs more from a single direction, the estimation error is larger, especially for the unbiased estimator.
In the case of the estimates obtained considering the spherical mean of the signal, the bias discussed here has to be considered as an additional one.

\subsection{Human scans}

\begin{figure}[h]
\centering
\includegraphics[width=0.97\textwidth]{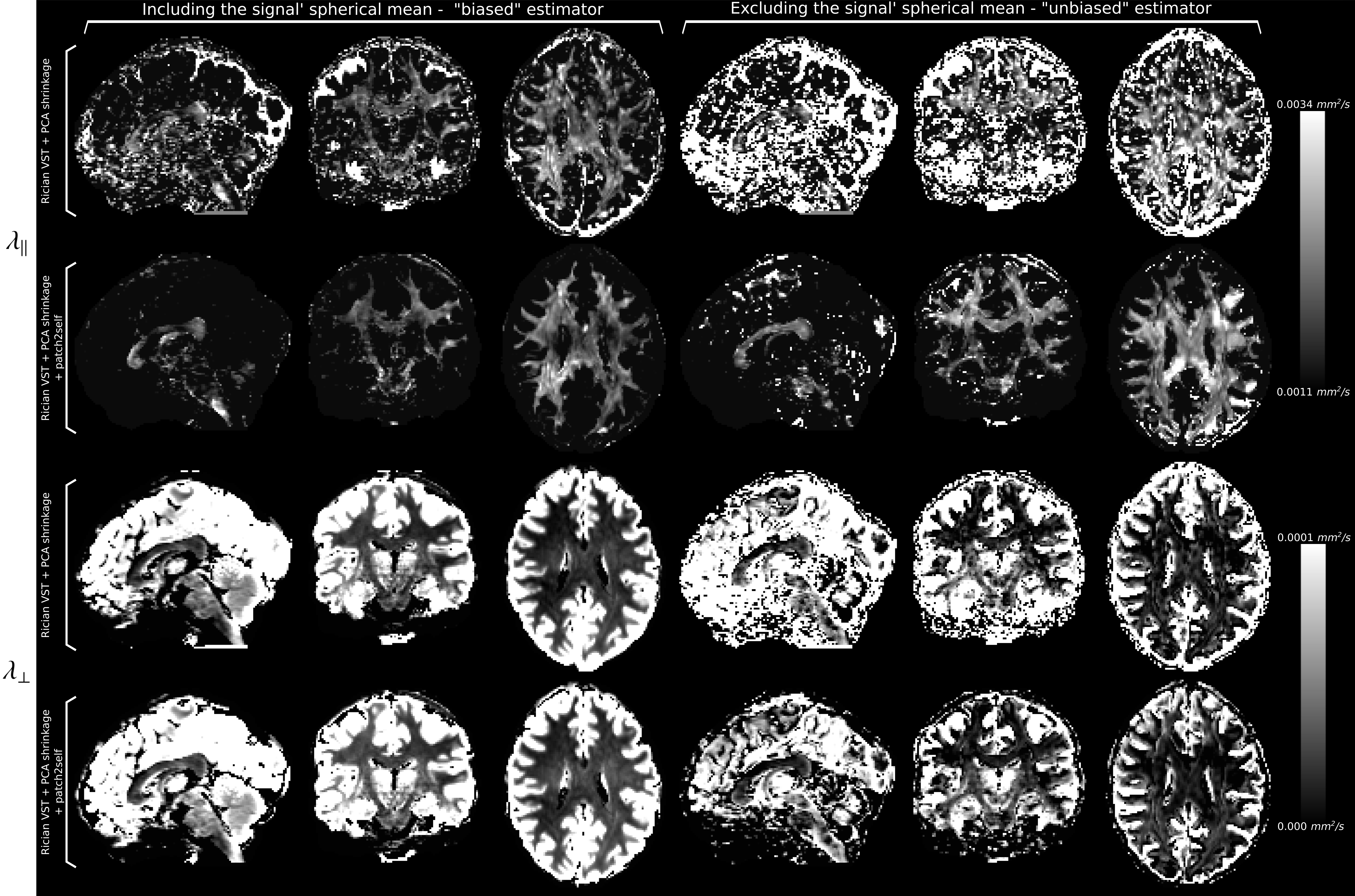}
\caption{Real data. Maps of parallel and perpendicular axonal diffusivities obtained for the subject number 1010 of the MGH-HCP Adult Diffusion dataset.
Maps were obtained using the variable projection method on the 5000 and 10000 $\textrm{s}/\textrm{mm}^2$ shells using the ``biased'' estimator that includes the spherical mean of the signal and the ``unbiased'' one on data processed by eventually including an additional restoration step (based on patch2self) after denoising using the Rician variance stabilization transformation (VST) and PCA shrinkage.}
\label{fig_single_subjects}
\end{figure}

\begin{figure}[h]
\centering
\includegraphics[width=0.97\textwidth]{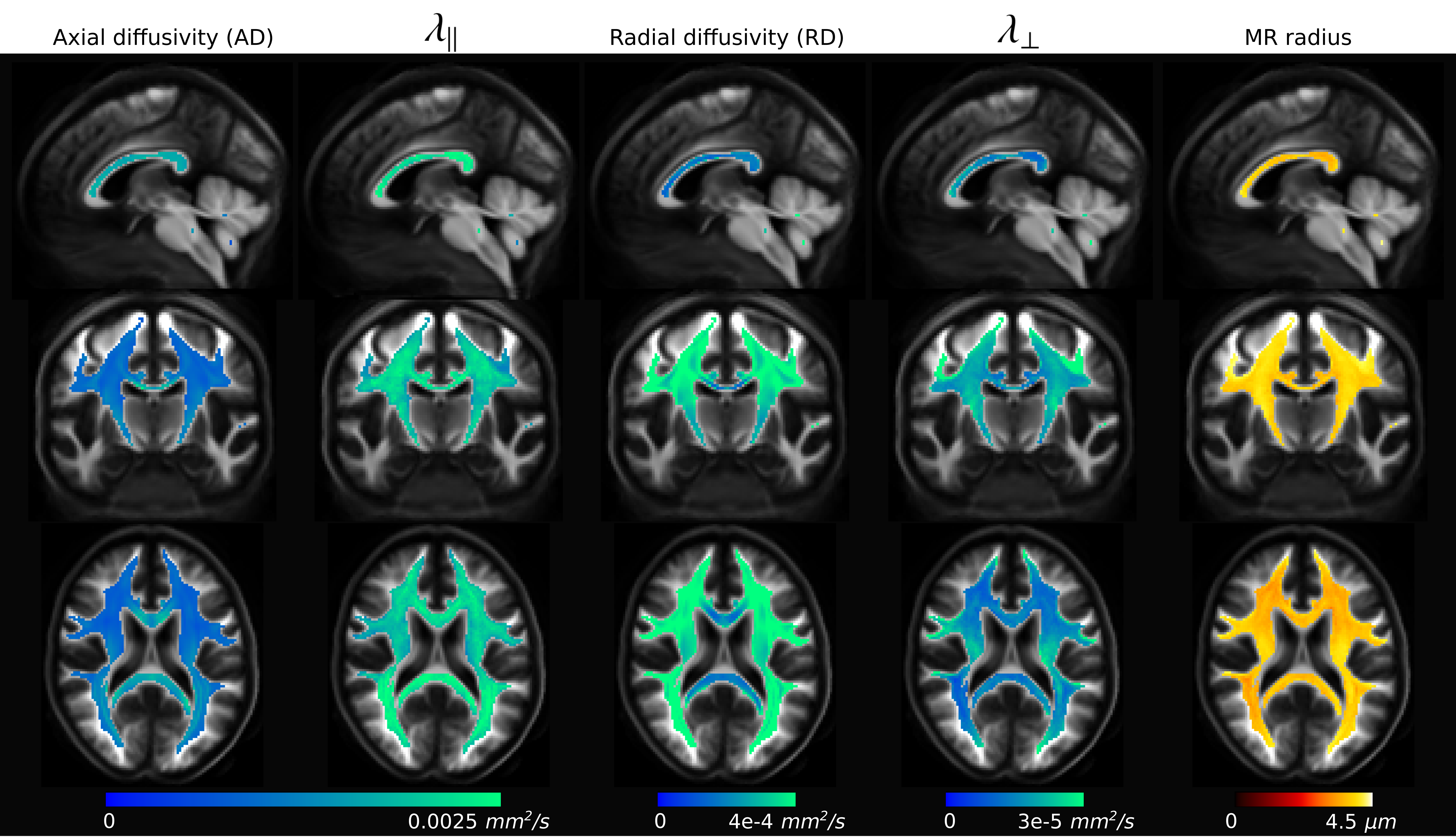}
\caption{Real data. Median values of axonal diffusivities (including the signal' spherical mean), DTI-based axial and radial diffusivity, and MR radius (calculated from the axonal diffusivities) obtained processing 34 subjects of the MGH-HCP Adult Diffusion database.}
\label{fig_template_medians}
\end{figure}

\begin{figure}[h]
\centering
\includegraphics[width=0.97\textwidth]{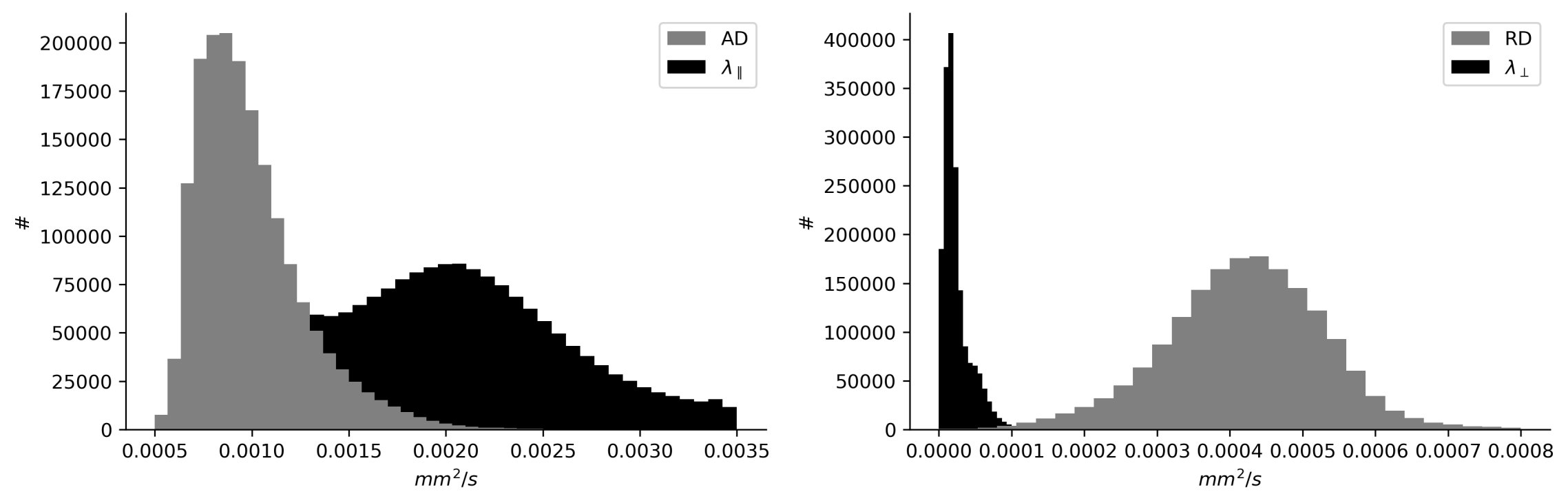}
\caption{Real data. Distribution of estimates of axonal parallel (left) an perpendicular (right) diffusivities in white matter obtained with the variable projection method (accounting for the spherical mean) and of axial and radial diffusivity calculated with DTI in the white matter region. The histograms report the number of voxels on the y axis and include results for 34 subjects of the MGH-HCP Adult Diffusion database.}
\label{fig_real_histograms}
\end{figure}

\begin{figure}[h]
\centering
\includegraphics[width=0.97\textwidth]{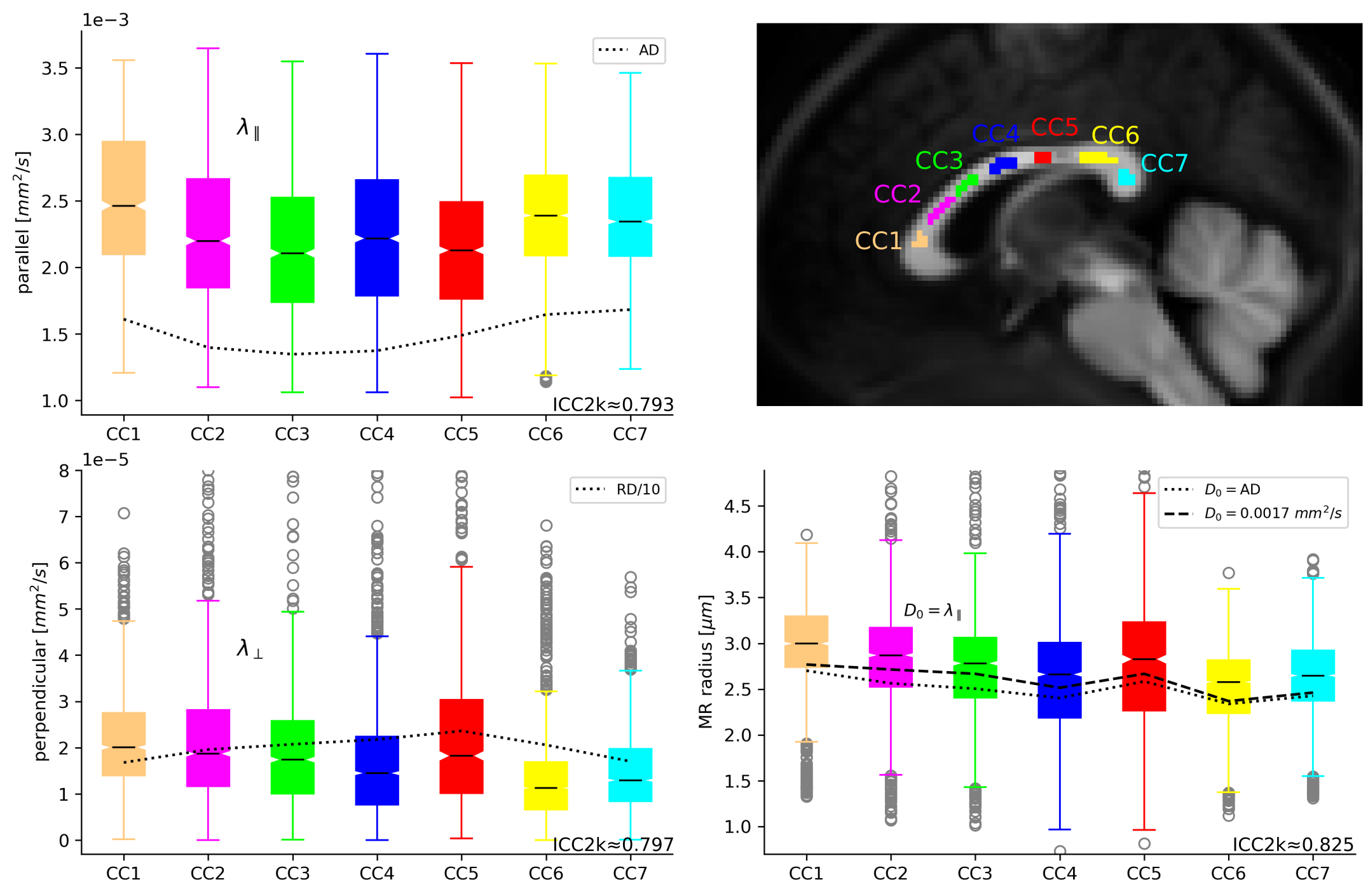}
\caption{Real data. Distributions of estimates of parallel, perpendicular diffusivity, and MR radius in different regions of the corpus callosum considering 34 subjects of the MGH-HCP Adult Diffusion dataset. Estimates were obtained using the variable projection estimator that includes the signal' spherical mean, after the data had been denoised. Intraclass correlation coefficients (2k) are reported. Values of axial (AD) and radial diffusivity (RD) - divided by 10 - from DTI are reported as well as the median values of the MR radius that would be obtained by fixing the intrinsic diffusivity to AD or by setting a specific value for it.}
\label{fig_cc_analysis}
\end{figure}

The method proved more robust -- less sensitive to noise -- in the presence of a significant amount of axons.
Therefore, in gray matter -- where the axonal fraction and the axonal signal-to-noise ratio are low -- the performance is poor.
Results are therefore discussed for a WM mask.
We report first the results where the VP method includes the spherical mean of the signal (``biased'' estimates) and adopts Laplace-Beltrami regularization.
To process the human data, we increased the regularization amount found with synthetic data to 25$\epsilon$ (biased estimator).
This is motivated by the fact that real data contains other artifacts that might not be entirely corrected by the preprocessing pipeline.
The maps of axonal diffusivities obtained for the 34 subjects look similar to those illustrated in fig.~\ref{fig_single_subjects}.
The median template values of the estimates are reported in fig.~\ref{fig_template_medians}.
The median value of the parallel and perpendicular axonal diffusivities over the whole WM mask are $\lambda_{\parallel}\approx0.0021\textrm{mm}^2/\textrm{s}$ and $\lambda_{\perp}\approx0.0000177\textrm{mm}^2/\textrm{s}$.
Fig.~\ref{fig_real_histograms} reports the histograms calculated across the 34 subjects.
We note that the distribution of the estimated $\lambda_{\parallel}$ increases towards the left-hand side of the mode due to the preference of the optimizer for values close to the imposed lower bound in some voxels which may be linked to the presence of noise.
In fig.~\ref{fig_cc_analysis} we report an analysis of the obtained estimates for various ROIs located the corpus callosum.
We note a correlation between AD and $\lambda_{\parallel}$ and a substantial anti-correlation between RD and $\lambda_{\perp}$.
The trend for the MR radius across the various corpus callosum regions is also reported, where the estimated value of $\lambda_{\parallel}$ was used as value for the intrinsic diffusivity, $D_0=\lambda_{\parallel}$.
Moreover, the median values corresponding to using the axial diffusivity, $D_0=$AD, and a constant value as used by \citet{fan2020axon}, $D_0=0.0017\textrm{mm}^2/\textrm{s}$, are also reported.
The statistical analysis in the corpus callosum based on the ICC2k (annotated for each plot in the figure) indicate that axonal diffusivities and MR radius estimates in the various regions of the corpus callosum are similar across subjects.

Given the sensitivity to noise of the unbiased estimates of the axonal diffusivities, we tried using \textit{patch2self} as an additional image restoration step of the DWIs that were already denoised.
The additional restoration step enabled the generation of visually appealing maps obtained with the unbiased estimator as illustrated in fig.~\ref{fig_single_subjects}.
While restoration indeed reduced the variability of estimates, experiments on synthetic data revealed that this comes at the cost of an additional \textit{restoration}-induced bias, compared to the non-restored DWIs, that becomes more relevant when including the spherical mean of the signal, as illustrated in supplementary fig.~S1.
The reduction of the variability of estimates is particularly evident for the unbiased estimates, which we illustrated for the corpus callosum in supplementary fig.~S2.
While the unbiased estimates may still be suboptimal in terms of bias, the reduced variability enabled obtaining visually appealing maps of the unbiased estimates where before the tissue contrast was hardly distinguishable.

\subsubsection{Partial volume bias}
\begin{figure}[h]
\centering
\includegraphics[width=0.97\textwidth]{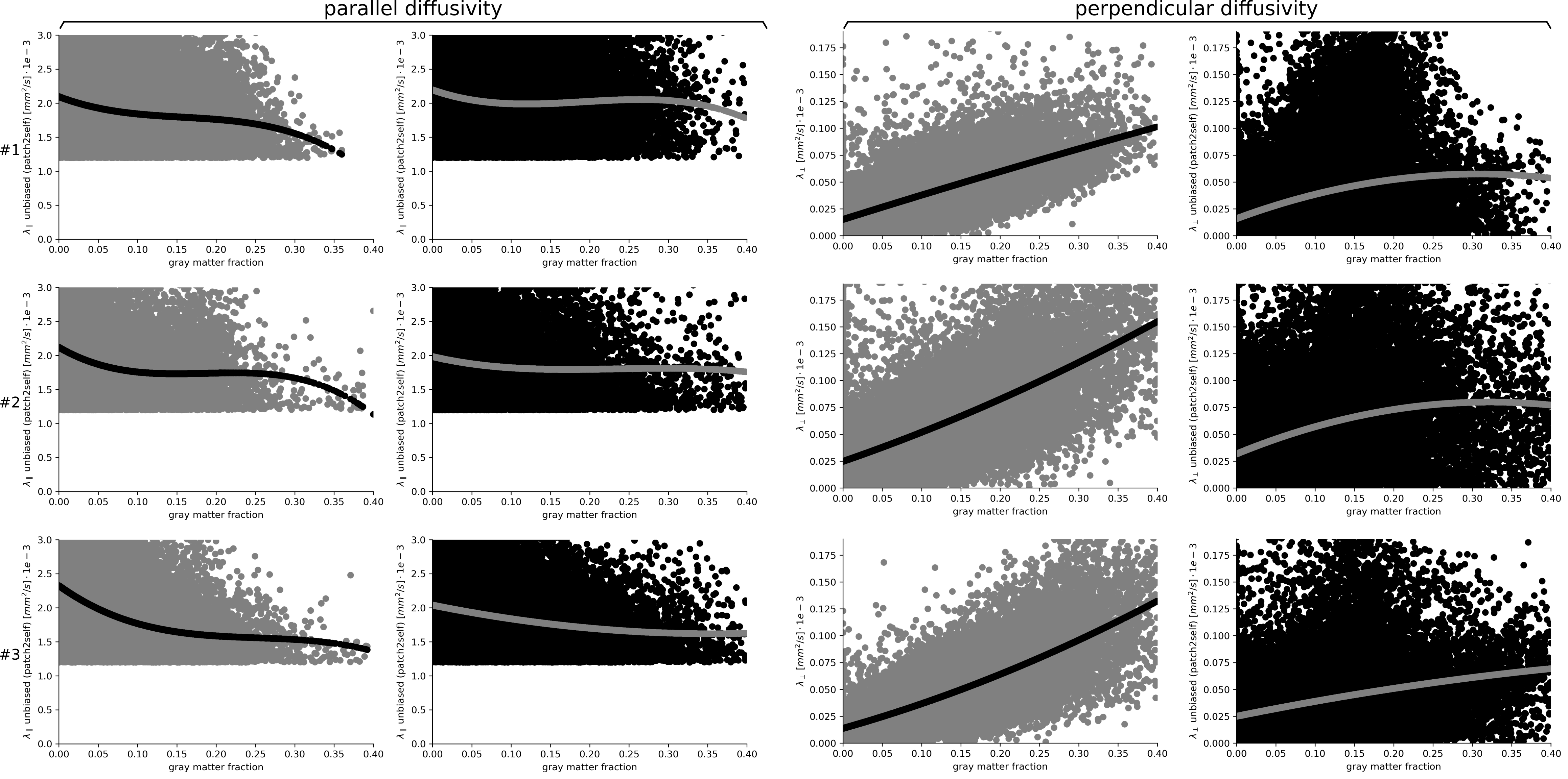}
\caption{Real data. Bias of axonal parallel (left pairs) and perpendicular (right pairs) diffusivities with respect to the gray matter fraction calculated for three selected subjects of the MGH-HCP Adult Diffusion database. In gray (with black regression line), the estimates obtained for the variable projection estimator that includes the signal' spherical mean, in black (with gray regression line) the estimates for the unbiased estimator after the data has been additionally restorated using patch2self.}
\label{fig_real_gm_bias}
\end{figure}

Results for partial volume with gray matter are illustrated in fig.~\ref{fig_real_gm_bias}.
These fundamentally confirm the findings obtained on synthetic data (fig.~\ref{fig_bias}).
The axonal diffusivity estimates obtained while accounting for the spherical mean show a clear correlation with the presence of partial volume with gray matter.
This correlation is reduced using the unbiased estimator, however a noticeable increase in the estimates' variability arises.
We finally observe that the scatter plots in fig.~\ref{fig_real_gm_bias} might be confounded by additional effects due to bias with respect to axonal dispersion or to the presence of multiple fiber bundles as described in section \ref{ssec_dispersion} and illustrated in fig.~\ref{fig_spherical_variance}.

%


\section{Discussion}
We proposed a method that allows estimating the axonal diffusivities and the MR radius from two single encoding PGSE shells acquired at strong diffusion weightings.
The use of single encoding PGSE data has several advantages.
First, it avoids mixing together data from different types of encoding where the diffusion weighting imparted to the tissue differently probes the underlying diffusion process, e.g., leading to different diffusion times.
Second, given a target diffusion weighting, the single encoding is that enabling to set the minimum echo time for the acquisition, which is advantageous in terms of SNR.
Finally, single encoding enables maximal angular resolution \citep{rensonnet2020signal} which is useful to resolve the directional information within a voxel.

The theoretical and experimental results demonstrate the feasibility of estimating axonal diffusivities using only two shells.
Notably, in the noise-free synthetic data the estimates are very close to the simulated values despite only modeling the axonal compartment and despite the fact that, along the direction of the axons, the parallel diffusivity simulated for the extra-axonal space was lower than that for the axonal space -- which implies that the signal parallel to the axonal direction is dominated by the extra-axonal contribution.
This therefore confirms that the proposed method, while sharing the advantage of rotational invariance of a powder averaging approach, enables the estimation of both axonal diffusivities.
Moreover, the synthetic results in the presence of noise (fig.~\ref{fig_synth_hist} right) suggest that the variable projection method leads to reduced variability in the estimates of perpendicular diffusivity compared to a power law approximation, i.e., Eq.~\ref{eq_powlaw_ratio_estimation}.
This is likely due to the fact that higher spherical harmonics zonal orders lead to higher sensitivity to both the parallel and the perpendicular axonal diffusivities as shown in fig.~\ref{fig_senitivities}.

The need for an unbiased estimator is highlighted by the results illustrated in fig.~\ref{fig_real_gm_bias}.
The proposed method theoretically enables estimating axonal diffusivities that are free from partial volume bias such as that due to the coexistence of gray and white matter tissue within a voxel.
This is enabled by the fact that zonal harmonic ratios with $L>0$ are insensitive to the spherical mean of the signal which includes contributions from not only from axons but also from other isotropic compartments \citep{zucchelli2017noddi,pizzolato2022axonal}.
In fact, as illustrated in fig.~\ref{fig_bias}, it is safe to state that modeling the axonal compartment at strong diffusion weightings without explicitly accounting for other compartments will always lead to partial volume bias, to a certain degree, unless strategies like the one proposed are adopted.
Although the theory allows for removal of the confounding factors related to the presence of residual partial volume at high b-values, the presence of residual noise limits its practical feasibility.
Indeed, the use of a strong diffusion weighting necessarily entails low SNRs.
Moreover, for magnitude DWIs, like in the case of the data used here, the noise deviates significantly from the Normal distribution as the SNR decreases, leading to well-known biases such as the Rician noise floor \citep{pizzolato2016noise,pizzolato2020adaptive}.
Whenever the removal of this bias is imperfect, for instance due to an incorrect estimation of the Gaussian noise variance, the values of the estimated axonal diffusivities can be biased.
This issue may be exacerbated with the data used for this study, where the presence of correlated noise limits the applicability of popular noise variance estimators based on principal component analysis.
In this work, we attempted to circumvent this issue by providing a different estimate of the noise variance through an alternative method.
Additionally, we attempted using a self-supervised machine learning method on the denoised data to further remove the residual noise.
While this approach enabled us to obtain visually appealing maps also with the unbiased estimator (fig.~\ref{fig_single_subjects}), and to retain to a certain degree the ability to remove partial volume bias (fig.~\ref{fig_real_gm_bias}), synthetic results highlighted the risks of incurring in additional bias (see supplementary figures).
The additional image restoration step, however, revealed that there may be ways to further reduce the variability of estimates thus enabling in the future the use of the unbiased estimator.

The presence of residual noise also leads to another form of bias when using the proposed method.
This is connected to the properties of the axonal orientation distribution function as summarized in fig.~\ref{fig_spherical_variance}.
In particular, the reduction of the higher orders spherical harmonics coefficients, due to a more isotropic orientation distribution function, paves the way for noise to affect the estimation of the axonal diffusivities.
Interestingly, the results in fig.~\ref{fig_spherical_variance} indicate that, although to a lesser extent, also the estimates obtained including the spherical mean of the signal suffer from this bias.
We point to the synthetic data analysis in \citet{pizzolato2022axonal} for a more specific analysis of the error where the effects of the number of fibers within a voxels is also analyzed, although in a different context.
It is possible to conclude that the results of the estimates we report are more reliable in areas of the white matter where the voxel contains only one fiber direction and that display a low amount of axonal dispersion, e.g., in the corpus callosum.
It is important to remind that this is not an intrinsic limitation of the proposed method, but rather an effect of residual noise.

The use of regularization like Tikhonov or Laplace-Beltrami is therefore recommended, particularly in real data where other forms of signal distortion may also be present.
This naturally comes at the cost of an additional, inevitable trade-off between bias and variance of the estimates.
Such a trade-off could be optimized using adaptive regularization criteria such as those discussed, for instance, for $T_2$ relaxometry \citep{canales2021comparison,canales2021revisiting}.
Moreover, in the future, the integration of necessary constraints \citep{haije2020enforcing} while fitting the signal could ensure that the optimization of Eq.~\ref{eq_optimization} can be performed while respecting the natural non-negativity of the signal.
The use of soft non-negativity constraints indeed revealed a prohibitive computational cost for the optimization. 
This type of constraints could reveal particularly useful when dealing with highly attenuated data, i.e., at strong diffusion weightings.

The effects of noise will likely be reduced when collecting data with larger voxel size and at lower b-value.
In fact, the highest b-value, 10000$\textrm{s/mm}^2$, can likely be reduced with great advantages in terms of effective SNR.

The values of the axonal diffusivities obtained with the proposed method -- when including the spherical mean of the signal for the estimates -- are in line with those obtained by \citet{ramanna2020triple}.
These differ substantially from the values obtained with DTI.
In particular, $\lambda_{\parallel}$ is larger than the axial diffusivity (even by as much as $60\%$ increase) and $\lambda_{\perp}$ is about one order of magnitude lower than the radial diffusivity.
The axial and radial diffusivities calculated from DTI include contributions from both the axonal and extra-axonal compartment and are affected by axonal orientation dispersion.
When considering the regions of interest in the corpus callosum -- where biases due to the presence of partial volume with gray matter or cerebrospinal fluid should be minimal -- it is then possible to conclude that the axonal parallel and perpendicular diffusivities must be larger and lower than the respective extra-axonal counterparts.
These findings are in line with those by \citet{kunz2018intra} in rats and by \citet{ramanna2020triple} in humans using prolate tensor encoding.
The values of MR radius are in line with the estimates obtained by \citet{veraart2020noninvasive} using higher b-values and multiple shells.
We note that the obtained estimates for the axonal diffusivities, illustrated for instance in fig.~\ref{fig_cc_analysis}, reflect an apparent diffusivity and are thus modulated by effects caused by the chosen timings, $\delta$ and $\Delta$, of the PGSE experiment.
Hence, the axonal parallel diffusivity does not directly correspond to the intrinsic diffusivity of the intra-axonal medium, $D_0$, and it rather is a (diffusion) time-dependent and lower estimate of it \citep{andersson2020axon} in the presence of undulation \citep{nilsson2012importance,brabec2020time} or of variations of the axonal diameter \citep{lee2019along,lee2020impact}.
The MR radius calculated using the estimated axonal parallel diffusivity -- although already being large compared to histological findings reported, for instance, by \citet{caminiti2009evolution} -- may nevertheless be underestimated compared to when the actual intrinsic diffusivity is used for the calculation.
In fact, the calculated MR radius depends on the chosen value for $D_0$ therefore, for the sake of comparison, in fig.~\ref{fig_cc_analysis} we also reported the median values that could be obtained when using a fixed value for the intrinsic diffusivity, as done by \citep{fan2020axon}, which leads to lower estimates.
Moreover, the weighting of the MR radius estimate towards the upper tail of the underlying diameter distribution moves it further from the true mean of said distribution, implying that the MR radius estimates must only be considered an indicative ``index''.

The final goal of microstructural characterization with diffusion-weighted MRI is the achievement of a robust quantification of the microstructural components that is reliably sensitive to changes caused by pathology.
In this scenario, the proposed method aims to isolate the axonal component and quantify its properties.
It can additionally be used as an initial step to characterize the remaining compartments with the use of a multi-compartment models \citep{alexander2010orientationally,pizzolato2018orientation,fan2020axon}.
The method can therefore be used to provide an informed initialization for the axonal diffusivities within a multi-compartmental model, and can be integrated into the optimization procedure.
We expect the practical feasibility of the method to broaden with the arrival of more powerful denoising methods and with slightly lower (high) b-values, such that the estimation of unbiased axonal diffusivities can be reliably deployed.

\section{Conclusion}
We have analyzed the problem of the estimation of the axonal diffusivities in the strong diffusion weighting regime, illustrating the underlying assumptions, possibilities, and limitations.
We have proposed a simplified linear equation for the estimation of the axonal perpendicular diffusivity.
Finally, we proposed a framework based on spherical harmonics for the variable projection estimation of the axonal diffusivities from two (single diffusion encoding) PGSE shells, and applied it to data from the MGH Adult Diffusion Human Connectome Project.
When estimating the axonal diffusivities from/including the directional spherical mean of the diffusion signal, we identified a source of bias due to residual partial volume with gray matter.
This bias can indeed be avoided using the proposed unbiased method although this makes the estimation more prone to variability due to residual noise after denoising.
Residual noise also determines the presence of a second bias which is related to the shape of the underlying orientation distribution of axons.
We reported the median values of the axonal parallel and perpendicular diffusivities and radii obtained while considering data from 34 subjects.
While the framework is solid and can theoretically produce unbiased estimates of axonal diffusivities, the presence of residual noise -- and potentially of other artifacts -- remains a major challenge given the low SNR of strongly diffusion-weighted data.
In fact, although the Tikhonov and Laplace-Beltrami regularization employed here were effective in obtaining more robust estimates, they too implement a trade-off between between bias and variance, which limits their effective applicability.
Future studies should therefore address these issues.


\section*{Acknowledgments}
Data were provided by the Human Connectome Project, WU-Minn Consortium (Principal Investigators: David Van Essen and Kamil Ugurbil; 1U54MH091657) funded by the 16 NIH Institutes and Centers that support the NIH Blueprint for Neuroscience Research; and by the McDonnell Center for Systems Neuroscience at Washington University.
Some image parts in the graphical abstract were kindly provided by Emmanuel Caruyer (\url{http://www.emmanuelcaruyer.com/q-space-sampling.php}).
This project has received funding from the European Union’s Horizon 2020 research and innovation programme under the Marie Skłodowska-Curie grant agreement No 754462 (Marco Pizzolato).
E.J. Canales-Rodr\'{i}guez was supported by the Swiss National Science Foundation (Ambizione grant PZ00P2\_185814).
Mariam Andersson was supported by the Capital Region of Denmark Research Foundation (grant number: A5657) (PI:Tim B. Dyrby).
\appendix

\section{Zonal harmonic functions}
\label{sec_zhc}
The even order functions $\Psi_l$ until $l=12$ are defined as
\begin{align*}
    \Psi_0(x) &= \phi_0(x) \\
    \Psi_2(x) &= [3\phi_2(x) - \phi_0(x)] / 2 \\
    \Psi_4(x) &= [35\phi_4(x) - 30\phi_2(x) + 3\phi_0(x)] / 8 \\
    \Psi_6(x) &= [231\phi_6(x) - 315\phi_4(x) + 105\phi_2(x) - 5\phi_0(x)] / 16 \\
    \Psi_8(x) &= [6435\phi_8(x) - 12012\phi_6(x) + 6930\phi_4(x) - 1260\phi_2(x) + 35\phi_0(x)] / 128 \\
    \Psi_{10}(x) &= [46189\phi_{10}(x) - 109395\phi_8(x) + 90090\phi_6(x) - 30030\phi_4(x) +3465\phi_2(x) -63\phi_0(x)] / 256 \\
    \Psi_{12}(x) &= [676039\phi_{12}(x) - 1939938\phi_{10}(x) + 2078505\phi_8(x) - 1021020\phi_6(x) + 225225\phi_4(x) - 18018\phi_2(x)\\
             \,&+ 231\phi_0(x)] / 1024
\end{align*}
where the functions $\phi_l$ are
\begin{align*}
    \phi_0(x) &= \frac{\sqrt{\pi}\textrm{erf}(x^{\frac{1}{2}})}{x^{\frac{1}{2}}} \\
    \phi_2(x) &= \frac{\sqrt{\pi}\textrm{erf}(x^{\frac{1}{2}}) - 2e^{-x}x^{\frac{1}{2}}} {2x^{\frac{3}{2}}} \\
    \phi_4(x) &= \frac{3\sqrt{\pi}\textrm{erf}(x^{\frac{1}{2}}) - 2e^{-x}x^{\frac{1}{2}}  (2x+3)} {4x^{\frac{5}{2}}} \\
    \phi_6(x) &= \frac{15\sqrt{\pi}\textrm{erf}(x^{\frac{1}{2}}) - 2e^{-x}x^{\frac{1}{2}}  (4x^2 + 10x + 15)} {8x^{\frac{7}{2}}} \\
    \phi_8(x) &= \frac{105\sqrt{\pi}\textrm{erf}(x^{\frac{1}{2}}) - 2e^{-x}x^{\frac{1}{2}}  (8x^3 + 28x^2 + 70x + 105)} {16x^{\frac{9}{2}}} \\
    \phi_{10}(x) &= \frac{945\sqrt{\pi}\textrm{erf}(x^{\frac{1}{2}}) - 2e^{-x}x^{\frac{1}{2}}  (16x^4 + 72x^3 + 252x^2 + 630x +945)} {32x^{\frac{11}{2}}} \\
    \phi_{12}(x) &= \frac{10395\sqrt{\pi}\textrm{erf}(x^{\frac{1}{2}}) - 2e^{-x}x^{\frac{1}{2}}  (32x^5 + 176x^4 + 792x^3 + 2772x^2 + 6930x + 10395)} {64x^{\frac{13}{2}}}
\end{align*}
refer to \citet{anderson2005measurement} or \citet{zucchelli2017noddi} for a description of the procedure required to derive the above relations.

\bibliographystyle{cas-model2-names}

\newpage
\bibliography{references}

\includepdf[pages={1,2}]{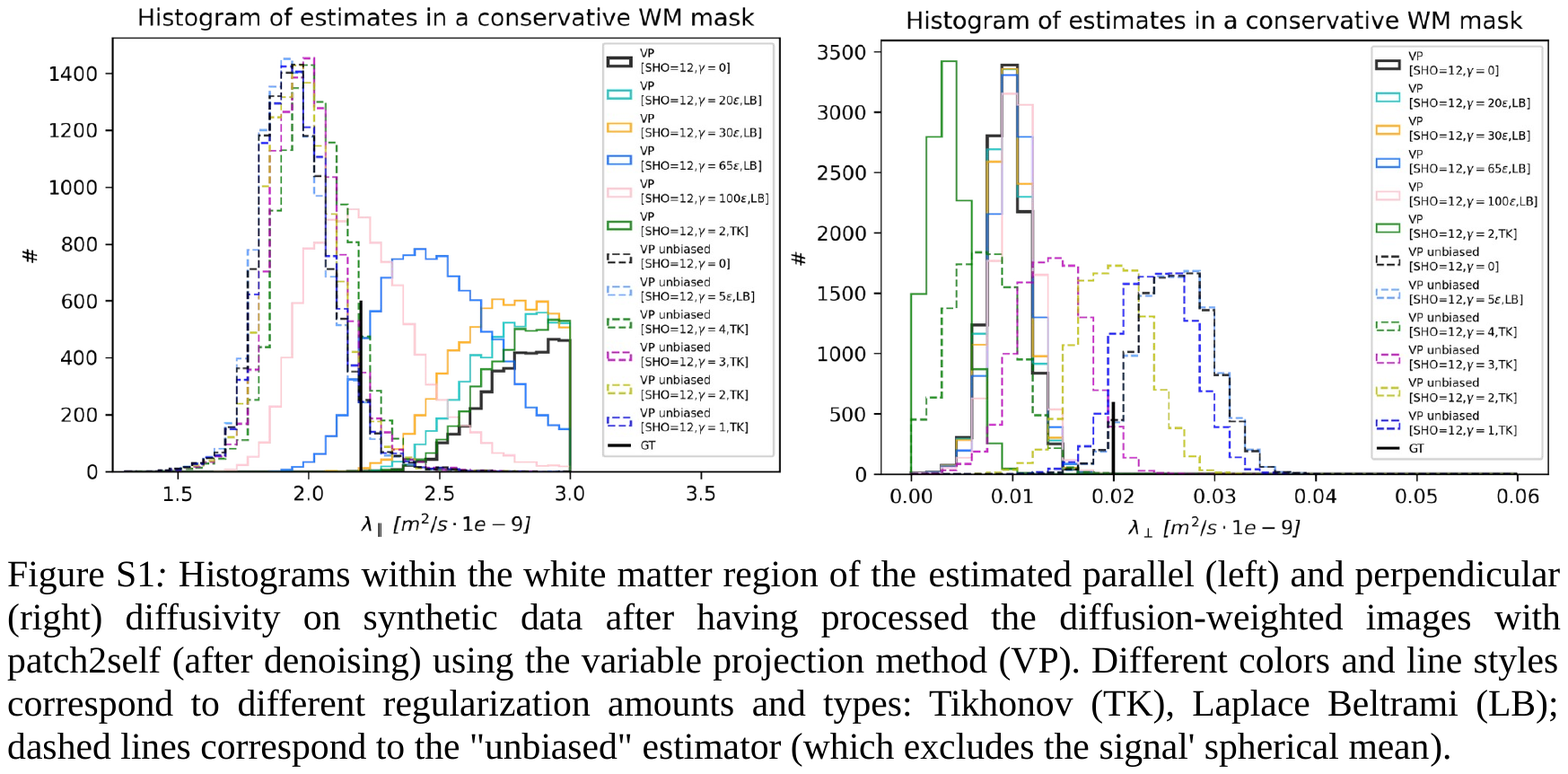}

\end{document}